{}

\documentclass[10pt,a4paper]{article}

\newif\ifpublic\publictrue

\ifpublic\else\usepackage{showkeys}\fi
\usepackage[bookmarks=true,hyperfigures=true,hidelinks,bookmarksnumbered,unicode]{hyperref}
\usepackage[nosort]{cite}
\usepackage[parsep]{collref}
\usepackage[english]{babel}
\usepackage[T1]{fontenc}
\usepackage[utf8]{inputenc}
\usepackage{lmodern}
\usepackage{amsfonts,amsbsy,dsfont,euscript,mathrsfs,fixmath}
\usepackage{amsmath,amssymb}
\usepackage{scalerel}
\usepackage{enumerate}
\usepackage{graphicx}
\usepackage{tikz,pgfplots}
\usetikzlibrary{arrows,calc,chains,decorations.pathmorphing,decorations.pathreplacing,fadings,matrix,positioning,shapes,shapes.geometric,shapes.symbols,trees}
\pgfplotsset{compat=1.9}

\def\showkeysrefformat#1{{\normalfont\tiny\ttfamily#1}}
\makeatletter
\def\SK@@ref#1>#2\SK@{{\@inlabelfalse\leavevmode\vbox to\z@{\vss\SK@refcolor\rlap{\vrule\raise .75em \hbox{\showkeysrefformat{#2}}}}}}
\makeatother

\usepackage[a4paper,text={160mm,247mm},centering]{geometry}
\allowdisplaybreaks[3]
\numberwithin{equation}{section}
\def\[{\begin{equation}\begin{aligned}}
\def\]{\end{aligned}\end{equation}}

\usepackage[font=small,labelfont=bf,width=0.85\textwidth]{caption}

\expandafter\def\expandafter\bfseries\expandafter{\bfseries\ifmmode\else\boldmath\fi}
\expandafter\def\expandafter\mdseries\expandafter{\mdseries\ifmmode\else\unboldmath\fi}
\expandafter\def\expandafter\normalfont\expandafter{\normalfont\ifmmode\else\unboldmath\fi}

\RequirePackage{verbatim}

\makeatletter
\newwrite\bibinl@out
\newenvironment{bibtex}[1][\jobname]{%
\immediate\openout\bibinl@out #1.bib%
\immediate\write\bibinl@out{\@percentchar generated from `\jobname' starting line \the\inputlineno^^J}%
\def\verbatim@processline{\immediate\write\bibinl@out{\the\verbatim@line}}%
\@bsphack\let\do\@makeother\dospecials\catcode`\^^M\active\verbatim@start%
}
{\immediate\closeout\bibinl@out\@esphack}
\makeatother

\let\barefrac=\frac
\renewcommand{\frac}[2]{\mathinner{\barefrac{#1}{#2}}}

\let\baresqrt=\sqrt
\makeatletter
\renewcommand{\sqrt}{\@ifnextchar[\@sqrt@space@a\@sqrt@space@b}
\def\@sqrt@space@a[#1]#2{\mathinner{\mathchoice{\mkern-3mu}{\mkern-3mu}{}{}\baresqrt[#1]{#2}}}
\def\@sqrt@space@b#1{\mathinner{\mathchoice{\mkern-3mu}{\mkern-3mu}{}{}\baresqrt{#1}}}
\makeatother

\makeatletter
\let\per@dot@old=\.
\def\.{\ifmmode\def\per@dot@sel{\mkern3mu}\else\def\per@dot@sel{\per@dot@old}\fi\per@dot@sel}
\makeatother

\let\barefootnote=\footnote
\renewcommand{\footnote}[1]{\barefootnote{#1\vspace{3pt}}}

\newcommand{\vfrac}[2]{\ifmmode\mathinner{\textstyle^{#1}\!/\!_{#2}}\else$^{#1}\!/\!_{#2}$\fi}

\DeclareMathOperator{\Tr}{Tr}

\newcommand{\Real}{\mathds{R}}
\newcommand{\Complex}{\mathds{C}}
\newcommand{\Integer}{\mathds{Z}}

\makeatletter
\newcommand*\bigcdot{\mathpalette\bigcdot@{.5}}
\newcommand*\bigcdot@[2]{\mathbin{\vcenter{\hbox{\scalebox{#2}{$\m@th#1\bullet$}}}}}
\makeatother

\newcommand{\id}{\mathrm{id}}

\newcommand{\alg}[1]{\mathfrak{#1}}
\newcommand{\grp}[1]{\mathrm{#1}}

\def\<{\big\langle}
\def\>{\big\rangle}

\newcommand{\bl}[2]{\<#1,#2\>}

\newcommand{\CP}{\Complex\mathbf{P}}

\newcommand{\Pexp}{\operatorname{Pexp}}

\newcommand{\Lag}{\mathcal{L}}
\newcommand{\Act}{\mathcal{S}}
\newcommand{\Ham}{\mathcal{H}}
\DeclareFontEncoding{LS1}{}{}
\DeclareFontSubstitution{LS1}{stix}{m}{n}
\DeclareSymbolFont{stixsymbols}{LS1}{stixscr}{m}{n}
\SetSymbolFont{stixsymbols}{bold}{LS1}{stixscr}{b}{n}
\DeclareMathSymbol{\kay}{\mathalpha}{stixsymbols}{"6B}
\DeclareMathSymbol{\hay}{\mathalpha}{stixsymbols}{"68}
\DeclareMathAlphabet{\mathdsl}{U}{bbm}{m}{sl}

\providecommand{\href}[2]{#2}

\makeatletter
\def\mr@ignsp#1 {\ifx\:#1\@empty\else #1\expandafter\mr@ignsp\fi}
\newcommand{\multiref}[1]{\begingroup%
\xdef\mr@no@sparg{\expandafter\mr@ignsp#1 \: }%
\def\mr@comma{}\def\mr@dash{-}%
\@for\mr@refs:=\mr@no@sparg\do{%
\ifx\mr@refs\mr@dash\def\mr@comma{}--\else%
\mr@comma\def\mr@comma{,}\ref{\mr@refs}\fi}%
\endgroup}
\renewcommand{\eqref}[1]{(\multiref{#1})}
\makeatother

\makeatletter
\newcommand{\namedref}[2]{\hyperref[#2]{#1~\ref*{#2}}}
\newcommand{\secref}{\@ifstar{\namedref{Section}}{\namedref{sec.}}}
\newcommand{\appref}{\@ifstar{\namedref{Appendix}}{\namedref{app.}}}
\newcommand{\tabref}{\@ifstar{\namedref{Table}}{\namedref{tab.}}}
\newcommand{\figref}{\@ifstar{\namedref{Figure}}{\namedref{fig.}}}
\makeatother

\let\oldbib=\thebibliography
\def\thebibliography{\phantomsection\addcontentsline{toc}{section}{\refname}\oldbib}

\let\oldtoc=\tableofcontents
\def\tableofcontents{\phantomsection\addcontentsline{toc}{section}{\contentsname}\oldtoc}

\providecommand{\hypersetup}[1]{}
\providecommand{\texorpdfstring}[2]{#1}
\hypersetup{plainpages=false}
\hypersetup{pdfpagemode=UseNone}
\hypersetup{bookmarksnumbered=true}
\hypersetup{pdfstartview=FitH}
\hypersetup{colorlinks=false}
\hypersetup{citebordercolor={1 1 1}}
\hypersetup{urlbordercolor={1 1 1}}
\hypersetup{linkbordercolor={1 1 1}}

\makeatletter
\let\@keywords\@empty
\let\@subject\@empty
\providecommand{\keywords}[1]{\gdef\@keywords{#1}}
\providecommand{\subject}[1]{\gdef\@subject{#1}}
\def\thetitle{\@title}
\def\theauthor{\@author}
\def\thesubject{\@subject}
\def\thedate{\@date}
\def\thekeywords{\@keywords}
\makeatother
\AtBeginDocument{
\hypersetup{pdftitle={\thetitle}}
\hypersetup{pdfauthor={\theauthor}}
\hypersetup{pdfsubject={\thesubject}}
\hypersetup{pdfkeywords={\thekeywords}}}

\newif\ifshownote
\shownotetrue

\ifpublic\shownotefalse\fi

\ifshownote

\ifpdf\else\RequirePackage[active]{srcltx}\fi
\RequirePackage{xcolor}
\RequirePackage{ulem}

\newcommand{\remark}[2][]{{\normalfont\sffamily\hspace{1ex}
\def\emph{\textsl}\def\textbullet{$\bullet$}
\def\tmparga{#1}
\def\tmpargb{BH}\ifx\tmparga\tmpargb\color[rgb]{0.7,0,0}\fi
\def\tmpargb{LTC}\ifx\tmparga\tmpargb\color[rgb]{0,0.7,0}\fi
\def\tmpargb{}\ifx\tmparga\tmpargb\normalfont\color{red}\fi
\def\tmpargb{}\ifx\tmparga\tmpargb\else \textbf{#1:}\fi
#2\hspace{1ex}}}

\else

\newcommand{\remark}[2][]{\ignorespaces}

\fi

\newcommand{\contract}{\mathbin{\raisebox{0.4ex}{$\lrcorner$}}}

\newcommand{\tn}[1]{\mathbf{\underline{#1}}}
\newcommand{\pd}{\partial}
\newcommand{\pdb}{\bar{\partial}}
\newcommand{\dr}{\mathrm{d}}
\newcommand{\rmi}{\mathrm{i}}

\newcommand{\cO}{\mathcal{O}}
\newcommand{\LieD}{\mathcal{L}}
\newcommand{\res}{\operatorname{res}}

\title{Integrable models from 4d holomorphic BF theory}
\author{Lewis T.\ Cole\texorpdfstring{$^{a,b}$}{} and Ben Hoare\texorpdfstring{$^a$}{}}

\begin{document}

\pdfbookmark[1]{Title Page}{title}
\thispagestyle{empty}


\vspace*{2cm}
\begin{center}
\begingroup\Large\bfseries\thetitle\par\endgroup
\vspace{1cm}

\begingroup\theauthor\par\endgroup

\vspace{1cm}

\begingroup\textit{
${}^a$\,Department of Mathematical Sciences,
\\Durham University, Durham DH1 3LE, UK
\\[0.1cm]${}^b$\,School of Mathematics and Maxwell Institute for Mathematical Sciences,
\\University of Edinburgh, Edinburgh EH9 3FD, UK}
\endgroup

\vspace{5mm}

\begingroup\ttfamily\small
lewis.cole@ed.ac.uk, ben.hoare@durham.ac.uk\par
\endgroup
\vspace{5mm}

\vfill

\textbf{Abstract}\vspace{5mm}

\begin{minipage}{12.5cm}\small
We show how to construct 2d field theories with holomorphic integrability from defect setups in 4d holomorphic BF.
In a simple example setup, we explicitly construct the 2d theory and perform an initial classical analysis.
Making use of the symmetries, we are able to write down an infinite family of solutions to the equations of motion.
Comparing with more typical integrable systems, we explain how 2d holomorphic integrability sits between the standard notions of integrability in one and two dimensions.
These 2d theories are designed as toy models for integrable theories in three and four dimensions, many of which can be understood as partially or totally holomorphic.
We comment on the implications for higher-dimensional integrability and aspects of quantization in the concluding remarks.
\end{minipage}

\vspace*{4cm}

\end{center}

\newpage
\tableofcontents
\vspace{2em}
\hrule
\vspace{2em}


\section{Introduction}\label{sec:intro}

The most well-known holomorphic integrable systems are 4d field theories whose equations of motion can be recast as the anti-self-dual Yang-Mills equations.
Such theories can be found from 6d holomorphic Chern-Simons (6d hCS) with twistor space as the underlying complex manifold~\cite{Costello:talk,Bittleston:2020hfv,Costello:2021bah}, and include the 4d Wess-Zumino-Witten model~\cite{Yang:1977zf,Pohlmeyer:1979ya,donaldson1985anti,Losev:1995cr,NekrasovThesis}, the 4d LMP model~\cite{Newman:1978ze,Leznov:1986mx,Leznov:1986up,Parkes:1992rz}, their gauged counterparts~\cite{Cole:2024sje} and deformed models~\cite{Cole:2023umd}.
On the other hand, there is a wealth of knowledge on 1d and 2d integrable theories, often expressed in the language of conserved charges and symmetry algebras~\cite{Lax:1968gpe,FaddeevTakhtajan,BabelonBernardTalon}.
These theories can be found from 3d mixed BF~\cite{Vicedo:2022mrm,Winstone:2023fpe,Caudrelier:2025xtx} and 4d mixed Chern-Simons~\cite{Costello:2019tri,Delduc:2019whp,Benini:2020skc} respectively, where the gauge theory is topological along the spacetime of the integrable theory in both cases.
Our goal is to better understand the relationship between holomorphic integrability and topological integrability by constructing:
\begin{itemize}
\item 2d holomorphic integrable systems starting from 4d holomorphic BF (4d hBF);
\item 2d integrable systems starting from 4d mixed Chern-Simons (4d CS);
\item 1d integrable systems starting from 3d mixed BF (3d BF);
\end{itemize}
and comparing how their classical integrability is manifested.
We will see that 2d holomorphic integrable systems sit between more traditional 1d and 2d integrable systems, as depicted in the following diagram:
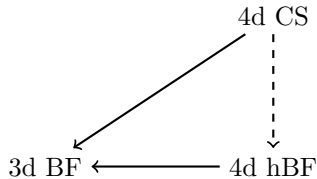
\begin{equation*}
\begin{tikzpicture}
\node (4mCS) at (6,1) {4d CS};
\node (3mBF) at (3,-1) {3d BF};
\node (4hBF) at (6,-1) {4d hBF};
\draw[thick,->] (4mCS)--(3mBF);
\draw[thick,->] (4hBF)--(3mBF);
\draw[thick,dashed,->] (4mCS)--(4hBF);
\end{tikzpicture}
\end{equation*}
The meaning of these theories and the arrows in this diagram will be explained throughout the paper.

\medskip

Many 2d integrable field theories (IFTs) can be found from 4d CS on $C \times \Sigma$ where $C$ is a Riemann surface and $\Sigma$ is 2d spacetime~\cite{Costello:2019tri,Delduc:2019whp,Benini:2020skc}, generalising earlier constructions in~\cite{NekrasovThesis,Costello:2013zra,Costello:2013sla,Costello:2017dso,Costello:2018gyb}, see \cite{Lacroix:2021iit} for a pedagogical review.
4d CS is holomorphic along $C$ and topological along $\Sigma$, hence the terminology mixed or holomorphic-topological.
Introducing defects localised at points in $C$ and specifying boundary and regularity conditions for the gauge field, the dynamics can be localised to $\Sigma$ and is described by a 2d IFT.
The emergence of a flat Lax connection
\begin{equation}
\dr L + L \wedge L = 0 ~,
\end{equation}
follows from the equations of 4d CS.
Moreover, the Poisson bracket of the Lax matrix is of Maillet form~\cite{Maillet:1985fn,Maillet:1985ek}, hence the path to conserved charges, both local involutive higher spin charges~\cite{Lacroix:2017isl}, and non-local Yangians and q-deformed algebras, is understood.
There is an analogous story relating 3d BF to 1d integrable models (IMs)~\cite{Vicedo:2022mrm,Winstone:2023fpe,Caudrelier:2025xtx}.
In this case, Lax's equation
\begin{equation}
\partial_t L = [M,L] ~,
\end{equation}
follows from the 3d BF equations and the Poisson bracket satisfies the standard Lax algebra.

The holomorphic integrability of the new 2d IFTs that we construct from 4d hBF is expected to be different in nature to either of these notions of integrability.
However, key observations that we exploit are that 4d hBF can be seen as a limit of 4d CS, and that its dimensional reduction is 3d BF~\cite{Winstone:2023fpe}.
By following this limit and reduction, we will be able to directly compare these different types of integrable systems.
Understanding the integrability of the new 2d IFTs may provide insights into the holomorphic integrability of the 4d IFTs found from 6d hCS \cite{Bittleston:2020hfv,Cole:2024sje,Cole:2023umd}.

\medskip

The holomorphic-topological gauge theories we have introduced thus far form part of a larger network of theories related by dimensional reductions and limits.
The original incarnation of this idea has a long history, starting with the anti-self-dual Yang-Mills equations in 4 dimensions and its reductions, see \cite{mason1996integrability} for a textbook introduction.
Many integrable systems can be found as reductions of these equations, for example: the Bogomolny equations in 3 dimensions; Hitchin's equations in 2 dimensions; and Nahm's equations in 1 dimension.
In fact, so many integrable systems arise is this manner, that the anti-self-dual Yang-Mills equations have been proposed as a `master' integrable system~\cite{Ward:1985gz}.

\begin{figure}
\begin{center}
\begin{tikzpicture}[scale=0.9]
\node (3CS) at (3,0) {\color{gray} 3d CS};
\node (4mCS) at (6,0) {4d CS};
\node (5mCS) at (9,0) {5d CS};
\node (6hCS) at (12,0) {6d hCS};
\node (2BF) at (0,-2) {\color{gray}2d BF};
\node (3mBF) at (3,-2) {3d BF};
\node (4hBF) at (6,-2) {4d hBF};
\node (1S) at (-3,-4) {\color{gray} 1d SB};
\node (2hS) at (0,-4) {2d hSB};
\draw[thick,->,gray] (4mCS)--(3CS);
\draw[thick,->] (6hCS)--(5mCS);
\draw[thick,->] (5mCS)--(4mCS);
\draw[thick,->] (5mCS)--(4hBF);
\draw[thick,->] (4mCS)--(3mBF);
\draw[thick,->] (4hBF)--(3mBF);
\draw[thick,->,gray] (3mBF)--(2BF);
\draw[thick,->,gray] (3CS)--(2BF);
\draw[thick,->,gray] (2BF)--(1S);
\draw[thick,->] (3mBF)--(2hS);
\draw[thick,->,gray] (2hS)--(1S);
\draw[thick,dashed,->] (4mCS)--(4hBF);
\draw[thick,dashed,->,gray] (3CS)--(3mBF);
\draw[thick,dashed,->,gray] (2BF)--(2hS);
\end{tikzpicture}
\end{center}
\caption{This figure depicts different possible dimensional reductions of 6d holomorphic Chern-Simons theory.
Schematically, horizontal arrows depict a reduction of a holomorphic direction, while diagonal arrows depict a reduction of a topological direction.
Dashed vertical arrows depict a limiting procedure in which one of the fields is taken to be large.
The black text and arrows can also be interpreted as a set of relations between IFTs in two fewer dimensions by picking out one of the holomorphic directions as the spectral plane.}\label{fig1}
\end{figure}
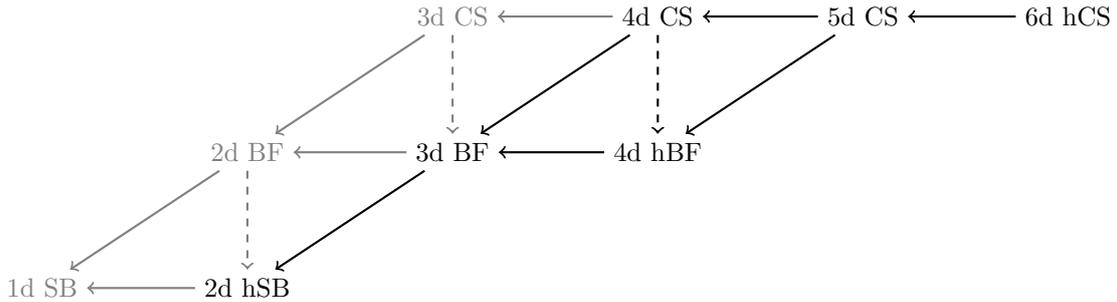

The more modern version of this hierarchy starts from 6d hCS, which can be used to construct 4d IFTs.
Schematically, 6d hCS only has holomorphic directions.
Picking one of these to reduce along, we find 5d mixed Chern-Simons (5d CS), with two holomorphic directions and one topological direction.
This theory can be used to construct 3d IFTs~\cite{Bittleston:2020hfv,Bittleston:2025gxr}.
When taking a reduction of 5d CS, we can choose whether to reduce along a holomorphic or a topological direction.
The former yields 4d CS and the latter gives rise to 4d hBF, two of the theories that we will study in this paper.
A further reduction, either topological or holomorphic, gives 3d mixed BF (3d BF) which can be used to construct 1d IMs~\cite{Vicedo:2022mrm,Winstone:2023fpe,Caudrelier:2025xtx}.
Finally, we can reduce along the remaining topological direction to yield a 2d holomorphic theory, which can either be understood as a gauged $\beta$-$\gamma$ system or holomorphic symplectic bosons (2d hSB).
This is the natural limit of this procedure if we would like to preserve at least one holomorphic direction that may interpreted as the spectral plane of an IFT or IM.
If all the directions are treated on an equal footing then we can go further, recovering purely topological theories including 3d CS, 2d BF and 1d SB.
This network of theories found as dimensional reductions of 6d hCS is depicted in \figref{fig1}.

The relationship between 4d CS and 3d CS first appeared in~\cite{Yamazaki:2019prm} where it was presented as a field-theoretic T-duality, and was generalised to include 5d CS and 6d hCS.
The reduction of 6d hCS to 5d CS and then to 4d CS together with the reductions of the associated IFTs was first presented in~\cite{Bittleston:2020hfv}.
The part of this network involving 5d CS, 4d CS, 4d hBF, and 3d BF first appeared in~\cite{Winstone:2023fpe}.
Different theories and relations between theories in this network are understood to varying degrees.

\medskip

The outline of this paper is as follows.
We start by introducing 4d hBF theory in~\secref{sec:4dhbf} and explaining its origin as a limit of 4d CS and its reduction to 3d BF.
We also explain how both 4d hBF and 4d CS can be found as reductions of 5d CS.
In~\secref{sec:defects}, we study the simplest 2d IFTs that originate from 4d hBF with defects, and relate them to 2d IFTs and 1d IMs coming from 4d CS and 3d BF respectively.
Throughout, we will highlight the new properties of 2d holomorphic integrability and compare them to the more standard notions of integrability in one and two dimensions.
Our analysis in this paper will be purely classical, though we comment on aspects of quantization in the concluding remarks in~\secref{sec:conc}.

\section{4d holomorphic BF theory}\label{sec:4dhbf}

\subsection{Defining 4d hBF theory}

Holomorphic BF theory was first introduced in~\cite{Popov:1999cq} and can be defined in any complex dimension.
Four-dimensional holomorphic BF theory is defined over a complex 2-manifold $M$ with coordinates $(z, w)$. The fundamental fields of this theory are a $(0,1)$-form gauge field $A = A_{\bar{z}} \dr \bar{z} + A_{\bar{w}} \dr \bar{w} \in \Omega^{0,1} (M, \alg{g})$ and an algebra-valued scalar field $b \in C^\infty (M, \alg{g})$. In the simplest case, the action is given by
\begin{equation}
\Act_{\text{hBF}_4} [b, A] = \frac{1}{2 \pi \rmi} \int_M \dr z \wedge \dr w \wedge \bl{b}{F(A)} ~,
\end{equation}
where $F(A) = \dr A + A \wedge A$ is the field strength of $A$ and we have denoted a non-degenerate bilinear form on the Lie algebra $\alg{g}$ by $\bl{\boldsymbol{\cdot}}{\boldsymbol{\cdot}} : \alg{g} \times \alg{g} \to \Complex$. For our purposes, it is sufficient to think of this bilinear form as the matrix trace in some representation of $\alg{g}$.
This action is invariant under gauge transformations of the form
\begin{equation}
A \mapsto g^{-1} A g + g^{-1} \dr g ~, \qquad
b \mapsto g^{-1} b g ~.
\end{equation}
These can be used to fix one of the components of $A$ to vanish, for example $A_{\bar{z}} = 0$, a gauge that we will consider shortly. The action is also invariant under shifting $A$ by any $(1,0)$-form $A \mapsto A + \sigma_z \dr z + \sigma_w \dr w$ and we will usually fix this symmetry by setting $A_z = 0$ and $A_w = 0$. Varying the action gives two equations of motion,
\begin{equation}
\delta \Act_{\text{hBF}_4} = \frac{1}{2 \pi \rmi} \int \dr z \wedge \dr w \wedge \bl{\delta b}{F(A)} + \frac{1}{2 \pi \rmi} \int \dr z \wedge \dr w \wedge \bl{\delta A}{\dr b + [A, b]} ~.
\end{equation}
The first of these tells us that the $(0,2)$-component of the field strength vanishes $F^{0,2}(A) = 0$, which implies that the connection $\overline{\nabla}_A = \pdb + A$ is holomorphic $[\overline{\nabla}_A, \overline{\nabla}_A] = 0$. The second equation of motion tells us that $b$ is holomorphic with respect to this connection $\overline{\nabla}_A b = 0$.

We can better understand the space of solutions to these equations by considering them in the gauge $A_{\bar{z}} = 0$. The first equation of motion becomes $\pd_{\bar{z}} A_{\bar{w}} = 0$, which implies that $A_{\bar{w}}$ must be a holomorphic function of $z$. This dramatically constrains the $z$-dependence of the gauge field. For example, if we also demand that $A$ tends to a finite value as $z \to \infty$ then the only solutions to $\pd_{\bar{z}} A_{\bar{w}} = 0$ are constant in the $z$-plane. The second equation of motion has two components, one of which $\pd_{\bar{z}} b = 0$ says that $b$ is also holomorphic in the $z$-plane, and the other $\pd_{\bar{w}} b + [A_{\bar{w}} , b] = 0$ is an equation in the $w$-plane that must hold for all values of $z$.

In this paper, we will be interested in constructing 2d integrable field theories living in the $w$-plane, and the complex variable $z$ will be treated as an auxiliary parameter that we refer to as the spectral parameter. The equations of motion $\pd_{\bar{z}} A_{\bar{w}} = 0$ and $\pd_{\bar{z}} b = 0$ will allow us to completely determine the $z$-dependence of both 4d fields, at which point the remaining equation $\pd_{\bar{w}} b + [A_{\bar{w}} , b] = 0$ can be treated as an equation in the $w$-plane. This final equation will be equivalent to the equations of motion of the 2d integrable field theory.
In effect, this is our definition of 2d holomorphic integrability -- a 2d field theory whose equations of motion can be recast in terms of the Lax equation $\pd_{\bar{w}} b + [A_{\bar{w}} , b] = 0$.

In order to get non-trivial 2d theories, we will need to introduce defects into the model, which will source singularities in the 4d fields. The on-shell field configurations will no longer be holomorphic functions of $z$, but rather meromorphic functions with singularities at the locations of the defects in the $z$-plane. The first method we consider for incorporating defects is to generalise the action by introducing a meromorphic $(2,0)$-form $\omega_2 = \varphi(z) \, \dr z \wedge \dr w$. The action of 4d hBF theory is then given by
\begin{equation}
\Act_{\text{hBF}_4} [b, A] = \frac{1}{2 \pi \rmi} \int \omega_2 \wedge \bl{b}{F(A)} ~.
\end{equation}
The meromorphic function $\varphi(z)$ will be treated as input data defining the theory and is not dynamical. While the bulk equations of motion for this theory remain unchanged, the variation of the action now includes a contribution of the form
\begin{equation}
\frac{1}{2 \pi \rmi} \int \dr \omega_2 \wedge \bl{\delta A}{b} ~.
\end{equation}
Since $\omega_2$ is meromorphic in the $z$-plane, the $3$-form $\dr \omega_2$ will be a distribution in the $z$-plane with support at the poles of $\omega_2$. This term in the variation will act as a boundary term and we impose boundary conditions on the fields such that it vanishes. These boundary conditions break the gauge invariance of the theory and give rise to local dynamics at the poles of $\omega_2$. It is convenient to introduce group-valued fields parametrising this broken gauge invariance that we will refer to as edge modes. These edge modes will become the dynamical field content of the 2d integrable theory.

In addition to poles, $\varphi(z)$ will also have zeroes at certain points in the $z$-plane. At these points, we can allow our field configurations to have singularities while still retaining a finite action. However, we need to ensure that the extent of the singularities in $A$ and $b$ do not conspire to overwhelm the zeroes of $\varphi(z)$. We will discuss precise field configurations in more detail when we turn to examples with explicit choices of $\varphi(z)$. Defects added to the theory by modifying the poles and zeroes of $\omega_2$ are known as ``disorder defects'' in the literature~\cite{Costello:2019tri}.

The second method we will consider for introducing defects into this theory is to explicitly add terms to the action at fixed points in the $z$-plane. For example, we will include terms of the form
\begin{equation}
\int_{\{z = \eta\}} \dr w \wedge \bl{\phi}{h^{-1} \overline{\nabla}_A h} \qquad \text{and} \qquad
\int_{\{z = \kappa\}} \dr^2 w \, \mathsf{P}(b) ~.
\end{equation}
In these expressions, $\phi \in C^\infty (M, \alg{g})$ and $h \in C^\infty(M, \grp{G})$ are degrees of freedom living on the defect, and $\mathsf{P} : \alg{g} \to \Complex$ is an ad-invariant polynomial on the algebra. The variations of these terms will source localised contributions to the equations of motion, which will in turn lead to singular on-shell field configurations. Such defects are known as ``order defects'' in the literature~\cite{Costello:2019tri}.

When considering order defects, it can be helpful to adopt the perspective presented in~\cite{Ashwinkumar:2023zbu}. Rather than thinking of them as arbitrary defect actions with undetermined dynamics (which may break integrability), we think of them as being built from the operators of the bulk theory itself. The bulk theory in~\cite{Ashwinkumar:2023zbu} is 4d CS, and order defects are interpreted as a thermodynamic limit of a mesh of Wilson lines. These Wilson lines can themselves be given various path integral representations, which involve introducing dynamical degrees of freedom living on the defects.

In our case, the gauge-invariant operators of 4d hBF theory are holomorphic Wilson lines of the partial connection $\overline{\nabla}_A$ and ad-invariant polynomials of the scalar field $b$. The first of our order defects looks like a coadjoint orbit realisation of a holomorphic Wilson line, and the second is an ad-invariant polynomial of $b$ integrated over the $w$-plane. While this perspective may be conceptually helpful, it will not play a practical role in our analysis.

\subsection{As a limit of 4d CS}

The type of 2d integrability inherited from 4d hBF theory is different from the usual type of 2d integrability inherited from 4d CS theory. The typical Lax formalism in 2d involves a $z$-dependent connection on spacetime which must be flat for all values of $z$ when the 2d equations of motion are satisfied. By comparison, the analogous structure we recover from 4d hBF theory is a $z$-dependent partial connection on spacetime together with a $z$-dependent section transforming in the adjoint representation. The integrability condition will be the statement that this section is holomorphic with respect to the partial connection for all values of $z$.

That the Lax on spacetime contains half as many components as we might expect is a consequence of 4d hBF theory being holomorphic along the $w$-plane rather than topological. In this sense, 4d hBF theory and the associated 2d integrable field theories are good toy models for the type of integrability appearing in 6d holomorphic CS theory and its associated 4d integrable field theories~\cite{Costello:2021bah,Bittleston:2020hfv,Cole:2023umd,Cole:2024sje}.

One of the main objectives of this paper is to understand how this holomorphic notion of integrability might be related to the more standard notion of integrability in 2d. We will explore this question by developing a concrete relationship between 4d hBF theory and 4d CS theory, as well as their associated 2d integrable theories. In particular, we will demonstrate that 4d hBF theory can be recovered as a limit of 4d CS theory, and that the same limit applies at the level of the 2d integrable theories. This provides an explicit map between the standard notion of 2d integrability and the alternative holomorphic notion.
At the level of the 4d theories, this limit has appeared in the literature in~\cite{Moosavian:2025nyv}.

Four-dimensional Chern-Simons theory~\cite{Costello:2013sla,Costello:2013zra,Costello:2017dso,Costello:2018gyb,Costello:2019tri} is defined over a 4-manifold $M_4 = C \times \Sigma$ which is the product of a complex 1-manifold $C$ with coordinate $z$ and a real 2-manifold $\Sigma$ with coordinates $x^a$ for $a \in \{1, 2\}$. The fundamental field is a $1$-form gauge field $A = A_{\bar{z}} \dr \bar{z} + A_a \dr x^a \in \Omega^1 (M_4, \alg{g})$ and the action is given by
\begin{equation}\label{eq:cs4}
\Act_{\text{CS}_4} [A] = \frac{k}{4 \pi \rmi} \int \omega_1 \wedge \bl{A}{\dr A + \frac{1}{3} [A, A]} ~.
\end{equation}
In this expression, we have introduced a meromorphic $(1,0)$-form $\omega_1 = \varphi(z) \, \dr z \in \Omega^{1,0}(C)$, which can have both poles and zeroes in the $z$-plane. This theory is known to describe a plethora of 2d integrable models~\cite{Costello:2019tri,Delduc:2019whp,Benini:2020skc}, with different models being related to different choices of $\omega_1$ and different boundary conditions on $A$ at the poles of $\omega_1$.

In order to recover 4d hBF theory as a limit of 4d CS theory, we need to introduce the complex coordinate $w$ as a function of $x^1$ and $x^2$. There is some subtlety in exactly how we achieve this and an associated choice\footnote{The two definitions of $w$ that we will consider in this paper are $w = x^1 + \rmi x^2$ and $w = x^1 + z \, x^2$.} that leads to different versions of 4d hBF theory. For the time being we will sidestep these issues and return to them later when we discuss explicit examples. Adopting the new coordinate, the 4d CS gauge field may be written as $A = A_{\bar{z}} \dr \bar{z} + A_{\bar{w}} \dr \bar{w} + A_w \dr w$. Unlike the 4d hBF gauge field, this $1$-form has a leg along $\dr w$ because $\omega_1$ does not saturate this direction. The component of $A$ along this direction will end up playing the role of the scalar field $b$ in the 4d hBF theory limit.

We can recover 4d hBF theory if we take the limit $k \to 0$ while also introducing a new parametrisation of the gauge field,
\begin{equation}
A = A^{0,1} + \frac{1}{k} b \, \dr w ~, \qquad
A^{0,1} = A_{\bar{z}} \dr \bar{z} + A_{\bar{w}} \dr \bar{w} ~.
\end{equation}
We have singled out the $\dr w$ direction and introduced a factor of $1/k$, which will diverge in the limit $k \to 0$. Since the action comes with an overall factor of $k$, only those terms that are linear in $b$ will survive in this limit. Substituting this parametrisation into the action gives
\begin{equation}
\Act_{\text{CS}_4} = \frac{1}{2\pi \rmi} \int \omega_1 \wedge \dr w \wedge \bl{b}{F(A^{0,1})}
+ \frac{1}{4 \pi \rmi} \int \dr \omega_1 \wedge \dr w \wedge \bl{b}{A^{0,1}}
+ \frac{k}{4 \pi \rmi} \int \omega_1 \wedge \bl{A^{0,1}}{\dr A^{0,1}} ~.
\end{equation}
In the limit $k \to 0$, the final term vanishes and the first two terms survive, landing on 4d hBF theory with a specific choice of boundary term,
\unskip\footnote{For the specific boundary conditions chosen in this paper, the boundary term in the action vanishes identically and does not affect the rest of our analysis. For this reason, we mostly drop it from the action in subsequent sections. This point is discussed more carefully in \secref{sec:minimal_setup}.}
\begin{equation}
\lim_{k \to 0} \Act_{\text{CS}_4} = \frac{1}{2 \pi \rmi} \int \omega_2 \wedge \bl{b}{F(A)} + \frac{1}{4 \pi \rmi} \int \dr \omega_2 \wedge \bl{b}{A} ~.
\end{equation}
The choice of $\varphi(z)$ does not change in this limit, and the $(2,0)$-form of 4d hBF theory is related to the $(1,0)$-form of 4d CS theory by $\omega_2 = \omega_1 \wedge \dr w$.

While this analysis describes the limit at the level of the action, we also need to provide boundary conditions on the 4d CS gauge field $A$ at the poles of $\omega_1$ to fully define the theory. We must then consider how these boundary conditions behave in the limit $k \to 0$ in order to deduce the resultant boundary conditions on the 4d hBF fields. Most boundary conditions will not be well-behaved in this limit. Indeed, any boundary conditions that make reference to the component $A_w$ will lead to a problematic $k\to0$ limit since this component diverges, $A_w \to \infty$. For this reason, the 4d CS setups we consider will involve boundary conditions that only depend on $A_{\bar{w}}$. This is both a highly restrictive choice and quite an atypical one when compared to the literature. We will therefore spend some time showing how we can find these setups as generalisations of the more typical relativistic setups in \secref{sec:defects4dCS}.

\subsection{Reductions to 3d BF}

We would also like to relate 4d hBF theory to another holomorphic-topological theory that has appeared in the literature. We will show that 3d BF theory, whose relationship to 1d integrable models has been studied in~\cite{Vicedo:2022mrm,Winstone:2023fpe,Caudrelier:2025xtx}, can be recovered as a 1d reduction of 4d hBF theory. This is analogous to the manner in which 5d CS and 4d CS are found as reductions of 6d hCS~\cite{Bittleston:2020hfv}.

Let us introduce a vector field $V$ on our 4-manifold satisfying $V \contract \dr w \neq 0$ and $V \contract \dr z = 0$. We would like to reduce 4d hBF theory along the flow generated by this vector field to recover a 3d theory on the quotient space. One can think of this as compactifying the integral curves of $V$ and then shrinking their radius to zero such that only the zero modes of the Kaluza-Klein tower survive. Practically speaking, we will implement this reduction in two steps. First, we will demand that our fields are invariant under the flow of $V$ by imposing the constraints $\LieD_V A = 0$ and $\LieD_V b = 0$. Then, we will contract $V$ into the $4$-form Lagrangian of 4d hBF theory to find a $3$-form that we identify as the Lagrangian of our 3d theory.

To simplify this calculation, it is helpful to fix the $(1,0)$-form shift symmetry $A \mapsto A + \sigma_z \dr z + \sigma_w \dr w$ in a different way. Rather than imposing $A_z = 0$ and $A_w = 0$, we will use this symmetry to impose the condition $V \contract A = 0$. This is possible because of the property $V \contract \dr w \neq 0$. In combination with the invariance constraint that we have already imposed, this ensures that $V \contract \bl{b}{F(A)} = 0$. Therefore, contracting $V$ into the 4-form Lagrangian of 4d hBF theory gives
\begin{equation}
V \contract \big( \omega_2 \wedge \bl{b}{F(A)} \big) = (V \contract \omega_2) \wedge \bl{b}{F(A)} ~.
\end{equation}
If we denote the $(1,0)$-form appearing on the right hand side by $\tilde{\omega}_1 = V \contract \omega_2$, then we can recognise the 3-form Lagrangian of 3d BF theory,
\begin{equation}\label{eq:3dbfaction}
\Act_{\text{BF}_3}[b,A] = \frac{1}{2 \pi \rmi} \int \tilde{\omega}_1 \wedge \bl{b}{F(A)} ~.
\end{equation}
This theory is known to describe a number of 1d integrable models, such as the finite Gaudin model~\cite{Vicedo:2022mrm,Winstone:2023fpe} and Hitchin's integrable system~\cite{Caudrelier:2025xtx}.

Let us note that we can similarly relate 4d CS theory~\eqref{eq:cs4} to 3d BF theory by a 1d reduction.
Proceeding in an analogous way, we introduce a vector field on our 4-manifold satisfying $V\contract \dr z = 0$.
We define the dual 1-form $\dr \rho$ such that $V\contract \dr \rho = 1$ and decompose the 4d CS gauge field as
\begin{equation}
A = A^\perp + b \, \dr \rho ~, \qquad b = V \contract A ~.
\end{equation}
Substituting this into the 4d CS action gives
\begin{equation}
\Act_{\text{CS}_4}[A] = \Act_{\text{CS}_4}[A^\perp] +\frac{1}{2\pi\rmi} \int \dr \rho \wedge \omega_1 \wedge \bl{b}{F(A^\perp)} + \frac{1}{4\pi\rmi} \int \dr \rho \wedge \dr \omega_1 \wedge \bl{b}{A^\perp} ~.
\end{equation}
To complete the reduction, we impose the constraint $\LieD_V A = 0$ and contract $V$ into the 4-form Lagrangian.
The only non-trivial contribution comes from the contraction of $V$ with the explicit factors of $\dr \rho$ yielding
\begin{equation}
\Act_{\text{BF}_3}[b,A^\perp] = \frac{1}{2\pi\rmi} \int \omega_1 \wedge \bl{b}{F(A^\perp)} + \frac{1}{4\pi\rmi} \int \dr \omega_1 \wedge \bl{b}{A^\perp} ~,
\end{equation}
which, upon renaming $\omega_1 \to \tilde\omega_1$ and $A^\perp \to A$, is the action of 3d BF theory~\eqref{eq:3dbfaction} with a specific choice of boundary term.

\subsection{Hamiltonian analysis of 4d hBF}\label{sec:hamiltonian4dhBF}

Let us review the Hamiltonian analysis of 4d hBF theory. This was first carried out in~\cite{Winstone:2023fpe} and closely follows the analysis of 3d BF theory~\cite{Vicedo:2022mrm}. We will focus on the bulk dynamics of the system (away from the poles of $\omega_2$) by neglecting any boundary terms that arise. This approach is already sufficient to learn something about 4d hBF theory: the 2d integrable models derived from this theory are guaranteed to be ultralocal, i.e., the Poisson bracket of $b$ with itself does not include the derivative of a delta function.

A 2d integrable field theory, together with its Lax, is said to be ultralocal if the Poisson bracket of the Lax matrix with itself contains only delta functions without spatial derivatives. By comparison, if the Poisson bracket of the Lax with itself contains derivatives of delta functions, it is said to be non-ultralocal. Theories with an ultralocal bracket often also have non-ultralocal brackets with the Lax matrices related by a gauge transformation, for example the sine-Gordon model~\cite{Vicedo:2017cge} and integrable sigma models with complex homogeneous~\cite{Bykov:2021dbk} and hermitian and para-complex coset \cite{Brodbeck:1999ib,Delduc:2019lpe} target spaces.

This distinction is important when it comes to quantizing 2d integrable field theories. For theories with an ultralocal bracket, the quantum inverse scattering method~\cite{Takhtajan:1979iv,Kulish:1979if,Sklyanin:1979pfu} is a well-established technique leading to an infinite tower of commuting charge operators, thereby manifesting the integrable structure of the quantum theory. However, the first-principles quantization of the integrable structure of truly non-ultralocal theories, including many 2d sigma models, presents ongoing challenges. One strategy~\cite{Faddeev:1985qu} is to modify the Poisson bracket and Hamiltonian to give a classically equivalent description of the model. When the new Poisson bracket is ultralocal, it is possible to employ the quantum inverse scattering method. However, in many cases it leads to a mildly non-ultralocal bracket~\cite{Semenov-Tian-Shansky:1995zdv}. The first-principles quantization of the resulting 2d integrable field theories remains an open problem~\cite{Delduc:2012qb,Delduc:2023exz}. It is therefore interesting that all of the integrable field theories derived from 4d hBF theory will be ultralocal.

Up to boundary terms, the action of 4d hBF theory can be written as
\begin{equation}
\Act_{\text{hBF}_4} = \int \dr z \wedge \dr \bar{z} \wedge \dr \bar{w} \wedge \dr w \, \Lag_{\text{hBF}_4} ~, \qquad
\Lag_{\text{hBF}_4} = -\frac{\varphi}{2 \pi \rmi} \bl{b}{\pd_{\bar{w}} A_{\bar{z}}} - \frac{\varphi}{2 \pi \rmi} \bl{A_{\bar{w}}}{\pd_{\bar{z}} b + [A_{\bar{z}}, b]} ~.
\end{equation}
In the Hamiltonian analysis, we treat $\bar{w}$ as the time direction,
\unskip\footnote{For example, we can analytically continue and treat $w$ and $\bar w$ as independent real variables.}
though one may also take the radial coordinate as time (as in~\cite{Winstone:2023fpe}) and the conclusion is unchanged.
With this choice, the action bears a striking resemblance to that of 3d BF theory, and as a result the Hamiltonian analyses of 3d BF and 4d hBF follow a similar derivation, albeit with the fields in the latter case depending on an extra variable. In some sense, this is an explanation of the fact that the 2d IFTs derived from 4d hBF are ultralocal: in the 1d IMs derived from 3d BF, there is simply no spatial direction along which to take a derivative of the delta function. Said differently, the derivative along the $w$-direction is manifestly absent from the action of 4d hBF. Let us present the Hamiltonian analysis to justify these remarks.

The conjugate momenta derived from the action are given by
\begin{equation}
P_b = 0 ~, \qquad
\Pi_{\bar{z}} = -\frac{\varphi}{2 \pi \rmi} b ~, \qquad
\Pi_{\bar{w}} = 0 ~.
\end{equation}
Moving to phase space, these definitions lead to three primary constraints on the system. Introducing the combination $\mathcal{C}_b = b + \frac{2 \pi \rmi}{\varphi} \Pi_{\bar{z}}$, we can write these constraints as
\begin{equation}
P_b \approx 0 ~, \qquad
\mathcal{C}_b \approx 0 ~, \qquad
\Pi_{\bar{w}} \approx 0 ~.
\end{equation}
The notation $\approx$ means that the two expressions agree on the constraint surface, i.e., when the constraints are imposed.
The first two of these constraints are second class as their Poisson bracket is invertible. This allows us to impose these constraints strongly by working on the reduced phase space defined by $P_b = 0$ and $\mathcal{C}_b = 0$. This does not affect the Poisson brackets of the other fields.

Applying a Legendre transform to the action, we find that the Hamiltonian density of this system can be written as
\begin{equation}
\Ham_{\text{hBF}_4} = \bl{A_{\bar{w}}}{\mu} ~, \qquad
\mu = \frac{\varphi}{2 \pi \rmi} \big( \pd_{\bar{z}} b + [A_{\bar{z}}, b] \big) ~.
\end{equation}
In this expression, $\mu$ is the moment map that generates gauge transformations in 4d hBF theory. Furthermore, if we compute the time evolution of the remaining constraint we find $\{ H_{\text{hBF}_4}, \Pi_{\bar{w}} \} = -\mu$ which implies that $\mu \approx 0$ is a secondary constraint. We can resolve this constraint by choosing a gauge fixing condition whose Poisson bracket with $\mu$ is invertible, and we will choose to fix $A_{\bar{z}} \approx 0$. Implementing this constraint requires working with the Dirac bracket on the reduced phase space, which we denote by $\{ \boldsymbol{\cdot}, \boldsymbol{\cdot} \}_{\ast}$. Here we will simply state the result and refer the reader to~\cite{Vicedo:2022mrm,Winstone:2023fpe} for further details.

One can show that the Dirac bracket on the reduced phase space of $\Pi_{\bar{z}}$ with itself is given by
\begin{equation}\label{eq:pbmomenta}
\{ \Pi_{\bar{z}}(z_1, w_1)_a , \Pi_{\bar{z}}(z_2, w_2)_b \}_{\ast} = \frac{1}{2 \pi \rmi} \frac{f_{ab}{}^c}{z_2 - z_1} \big( \Pi_{\bar{z}}(z_1, w_1)_c - \Pi_{\bar{z}}(z_2, w_2)_c \big) \, \delta(w_1 - w_2) ~,
\end{equation}
where $f_{ab}{}^c$ are the structure constants of the Lie algebra $\alg{g}$.
This takes the form of the Sklyanin bracket~\cite{Sklyanin:1982tf} and is the standard form of the Lax algebra for an ultralocal 2d integrable theory with the R-matrix
\begin{equation}
\mathcal{R}_{\tn{12}}(z_1,z_2) = \frac{C_\tn{12}}{z_2 - z_1} ~,
\end{equation}
where $C_{\tn{12}}$ is the quadratic Casimir.
In terms of the fundamental field $b$, the Poisson bracket~\eqref{eq:pbmomenta} may be written as
\begin{equation}\label{eq:ultralocal}
\{ b(z_1, w_1)_a , b(z_2, w_2)_b \}_{\ast} = \frac{f_{ab}{}^c}{z_1 - z_2} \bigg( \frac{b(z_1, w_1)_c}{\varphi(z_2)} - \frac{b(z_2, w_2)_c}{\varphi(z_1)} \bigg) \, \delta(w_1 - w_2) ~.
\end{equation}

To interpret this result, we can make comparisons with 4d CS and 3d BF. When compared with 4d CS, it appears that 4d hBF only has access to a smaller class of 2d integrable models, the dynamics of which are always ultralocal. On the other hand, it is also possible to recover ultralocal models from 4d CS, for example~\cite{Costello:2019tri,Caudrelier:2020xtn,Fukushima:2020tqv,Ashwinkumar:2023zbu}, so perhaps 4d hBF should be thought of as a special sector of 4d CS.
We can understand the ultralocal Poisson bracket~\eqref{eq:ultralocal} as the $k\to0$ limit of the non-ultralocal Poisson bracket of 4d CS with disorder defects~\cite{Vicedo:2019dej}
\begin{equation}\begin{split}\label{eq:nonultralocal}
\{ A_w(z_1,w_1)_a , A_w(z_2,w_2)_b \}_{\ast} & = \frac{f_{ab}{}^c}{z_2 - z_1} \bigg( \frac{A_w(z_1, w_1)_c}{k\varphi(z_2)} - \frac{A_w(z_2, w_2)_c}{k\varphi(z_1)} \bigg) \, \delta(w_1 - w_2)
\\ & \quad - \frac{\kappa_{ab}}{k}\bigg(\frac{\varphi(z_2)^{-1} - \varphi(z_1)^{-1}}{z_2-z_1}\bigg) \delta'(w_1-w_2) ~,
\end{split}\end{equation}
where $\kappa_{ab}$ are the components of the Killing form of the Lie algebra $\alg{g}$.
This takes the form of the Maillet bracket~\cite{Maillet:1985fn,Maillet:1985ek} with the R-matrix
\begin{equation}
\mathcal{R}_{\tn{12}}(z_1,z_2) = \frac{C_\tn{12}}{z_2 - z_1}\frac{1}{k\varphi(z_2)} ~.
\end{equation}
Setting $A_w = k^{-1} b$ in~\eqref{eq:nonultralocal}, the non-ultralocal term is subleading in the $k\to0$ limit and we recover~\eqref{eq:ultralocal}.

When compared with 3d BF, there are striking similarities in the Hamiltonian analysis. Since 3d BF describes 1d integrable models, it is not possible for their Poisson brackets to contain spatial derivatives, and in this loose sense they are all ultralocal. This leads us to the perspective that 4d hBF is describing a 2d theory in between a 1d integrable theory and a 2d integrable theory, such as a holomorphic integrable theory in one complex dimension.

Let us consider an arbitrary 3d BF theory setup and its associated 1d integrable theory. If we promote the time variable $t \in \Real$ to a complex variable $\bar{w} \in \Complex$, we can consider the equation of motion as an equation in one complex dimension. These are precisely the types of theories that we might expect to recover from 4d hBF theory. Indeed, derivatives along the $w$-direction cannot appear in the Hamiltonian analysis because they are absent from the action itself. This argument indicates that we can recover holomorphic integrable theories by complexifying the spacetime of a standard 1d IM.

The holomorphic integrable systems we have just described are not particularly interesting since, at least in some sense, they are still 1d rather the 2d systems. For example, the 2d action will only contain $\bar{w}$-derivatives of the fundamental fields. To work around this property, let us consider the map between the $(w, \bar{w})$ coordinates and the real coordinates of 2d spacetime. We will argue that introducing a $z$-dependence into this map can produce 2d integrable theories from 4d hBF theory whose actions contain derivatives in multiple different directions. This is the same mechanism by which anti-self-dual Yang-Mills is recovered from 6d hBF theory whose underlying complex manifold is twistor space~\cite{Witten:2003nn,Mason:2005zm}.

By introducing a $z$-dependence into this map, the interpretation of the Poisson bracket~\eqref{eq:ultralocal} as equal-time becomes more opaque.
This is because the coordinate that should be set equal will depend on 2d spacetime and the $\CP^1$ coordinate $z$.
\unskip\footnote{We thank one of the anonymous referees for pointing this out.}
It would be interesting to investigate both the more standard approach of choosing one of the coordinates on 2d spacetime as the time coordinate and the effect this has on the Poisson bracket, as well as the implications of the Poisson bracket~\eqref{eq:ultralocal} after introducing the dependence on $z$.
While we leave a detailed study for the future, we do carry out a more canonical Hamiltonian analysis for the simplest 2d IFT that originates from 4d hBF with defects in~\secref{sec:analysis} and again find an ultralocal structure.

\subsection{Underlying complex manifold}

By default, it appears that the 2d integrable models derived from 4d hBF theory will only contain $\bar{w}$-derivatives in their actions and equations of motion. In order to work around this property, we will introduce a slight generalisation of the 4d hBF theory we have discussed so far. Rather than defining 4d hBF theory over a complex 2-manifold with coordinates $(z, w)$, we will define it over the product of a complex 1-manifold with coordinate $z$ and a real 2-manifold with coordinates $x^a$ for $a \in \{1, 2\}$. We will then provide a map $(z, x^a) \mapsto (z, w)$ and pullback the meromorphic $(2,0)$-form $\omega_2$ along this map. In other words, we do not identify $(w, \bar{w})$ as coordinates on 2d spacetime, but rather as coordinates on an underlying complex manifold.

Let us choose the complex manifold parametrised by $z$ to be $\CP^1 = \Complex \cup \{ \infty \}$ and the real manifold parametrised by $x^a$ to be $\Real^2$. We would like to compare two different maps from this space to a complex 2-manifold. These maps are given by
\begin{equation}
p_0 : (z, x^a) \mapsto (z, w = x^1 + \rmi x^2) ~, \qquad
p_1 : (z, x^a) \mapsto (z, w = x^1 + z \, x^2) ~.
\end{equation}
The first map corresponds to equipping the 2d spacetime $\Real^2$ with a complex structure, and the underlying complex manifold is the product manifold $\CP^1 \times \Complex$. We can also think of this as a trivial rank 1 vector bundle over $\CP^1$ where $w$ is the coordinate along the fibre and $z$ is the coordinate on the base. The second map\footnote{In homogeneous coordinates $[\lambda_1, \lambda_2] \in \CP^1$ the second map reads $p_1 : (\lambda_a , x^a) \mapsto (\lambda_a, w = \lambda_a x^a)$.} is more complicated as it mixes the spacetime coordinates $x^a$ with the spectral parameter $z$. We can understand the underlying complex manifold by considering another rank 1 vector bundle over $\CP^1$, this time with a non-trivial transition function.

Consider the bundle $\cO(1) \to \CP^1$ whose base manifold is $\CP^1$ and whose fibres are $\Complex$. The non-trivial data defining this bundle is the transition function as we move between the two patches covering $\CP^1$. If the southern patch is parametrised by the coordinate $z$, then the coordinate on the northern patch is defined by $\tilde{z} = 1/z$. Similarly, as we move between these patches the fibre coordinate transforms from $w$ to $\tilde{w} = w/z$. Holomorphic sections of this bundle then take the general form $w = x^1 + z \, x^2$ where $x^a$ are complex parameters. 
Given a fixed element $x^a \in \Real^2$, the image of $p_1$ defines a holomorphic section of $\cO(1)$ for which the parameters $x^a$ are real, which is not the most general holomorphic section.
This is the underlying complex manifold that we will consider for the remainder of this paper.

An additional subtlety when working with the underlying complex manifold $\cO(1)$ is that $(w, \bar{w})$ are no longer good coordinates on 2d spacetime. The vector fields $\{ \pd_w , \pd_{\bar{w}} \}$ do not span the tangent space of $\Real^2$ at every point in $\CP^1 \times \Real^2$ as they fail to be independent whenever $z$ is real. We will remedy this by adopting a new coordinate $\hat{w} = (x^2 - \bar{z} \, x^1)/(1 + z \bar{z})$ that\footnote{In homogeneous coordinates, we can write $\hat{w} = \hat{\lambda}_a x^a / \lambda^b \hat{\lambda}_b$ where $\hat{\lambda}_a = (- \bar{\lambda}_2, \bar{\lambda}_1)$ and $\lambda^a = \lambda_b \varepsilon^{ba}$.} is independent of $w$ for all values of $z$. Fortunately, we can simply replace $\bar{w}$ with $\hat{w}$ in all of our previous analysis since we did not rely on the reality conditions on these coordinates at any point. Let us revisit the definition of 4d hBF theory so that we can expand upon this point.

We have defined 4d hBF theory over the manifold $\CP^1 \times \Real^2$ with coordinates $z \in \CP^1$ and $x^a \in \Real^2$ for $a \in \{1, 2\}$. Recall, the action of 4d hBF theory is given by
\begin{equation}
\Act_{\text{hBF}_4} [b, A] = \frac{1}{2 \pi \rmi} \int \omega_2 \wedge \bl{b}{F(A)} ~.
\end{equation}
It will be convenient for us to take $b$ to be a section of $\cO(-1)$ meaning that it transforms as $b \mapsto z \, b$ as we move from the southern to the northern patch on $\CP^1$. The action is only well-defined if the integrand is weight zero, so we take the meromorphic $(2,0)$-form $\omega_2$ to be valued in $\cO(1)$ to compensate.

The meromorphic $(2,0)$-form takes the form $\omega_2 = \varphi(z) \, \dr z \wedge \dr w$ where $w = x^1 + z \, x^2$ should be understood as a function of $z$ and $x^a$. It is useful to work in a basis of $1$-forms on $\CP^1 \times \Real^2$ that are adapted to $\omega_2$, hence we write the $1$-form gauge field as
\begin{equation}
A = A_{\bar{z}} \dr \bar{z} + A_{\hat{w}} e^{\hat{w}} ~, \qquad
e^{\hat{w}} = \frac{\dr x^2 - \bar{z} \, \dr x^1}{1 + z \bar{z}} ~,
\end{equation}
where $e^{\hat{w}}$ can be thought of as a frame along the $\hat{w}$ direction.
In this basis, we have $\pd_{\bar{z}} \contract A = A_{\bar{z}}$ where the vector field $\pd_{\bar{z}}$ is defined by holding $x^a$ fixed rather than holding $(w, \hat{w})$ fixed. In this expression, the combination $A_{\hat{w}} e^{\hat{w}}$ is playing the role that $A_{\bar{w}} \dr \bar{w}$ played in our previous notation. Since our analysis neither relied on the explicit form of $\bar{w}$ nor the fact that it was related to $w$ under complex conjugation, we can substitute the new coordinate $\hat{w}$ into the previous formulae and arrive at the same conclusions.

The advantage of working with this underlying complex manifold is that the complementary direction $\hat{w}$ is now a $z$-dependent function of the spacetime coordinates. The derivative in this direction can be written\footnote{In homogeneous coordinates, we can write $\pd_{\hat{w}} = \lambda^a \pd_a$.} as $\pd_{\hat{w}} = \pd_2 - z \, \pd_1$ (where $\pd_a = \pd/\pd x^a$), which means that, if we evaluate this derivative at different points in the $z$-plane, we recover different directional derivatives along spacetime. In this sense, we expect the 2d integrable theories to be more complicated (and interesting) than the direct complexification of a 1d IM.

\subsection{Reductions of 5d CS theory}

Before we proceed to discuss some examples of defect setups in the aforementioned theories, let us explain how 5d CS can reduce to either 4d CS or 4d hBF, as depicted in \figref{fig1}. The theory that we land on depends on whether we reduce along a holomorphic direction, in which case we find 4d CS, or a topological direction, in which case we find 4d hBF. Let us substantiate these claims.

We define 5d CS theory over the manifold $\CP^1 \times \Real^3$ by the action functional
\begin{equation}
\Act_{\text{CS}_5} [A] = \frac{1}{4 \pi \rmi} \int \omega_2 \wedge \bl{A}{\dr A + \frac{1}{3} [A, A]} ~.
\end{equation}
Just like 4d hBF theory, this theory depends on some additional data, namely a map from $\CP^1 \times \Real^3$ to a complex 2-manifold, which we denote by $\mathsf{E}$, and a meromorphic $(2,0)$-form on that complex 2-manifold. We then pullback this $(2,0)$-form via the map $p : \CP^1 \times \Real^3 \to \mathsf{E}$ and the result is $\omega_2$ which appears in the action.

For example, we can take $\mathsf{E}$ to be a rank-1 holomorphic vector bundle over $\CP^1$, just like the bundle $\cO(1) \to \CP^1$ that we used in our construction of 4d hBF theory. Any rank-1 holomorphic vector bundle over $\CP^1$ takes the form $\cO(n) \to \CP^1$ for some integer $n \in \Integer$. These bundles all have $\Complex$ fibres with the base manifold $\CP^1$ and they are distinguished by their transition functions when moving between the southern and northern patches of $\CP^1$.

In the literature, 5d CS theory and its non-commutative counterpart have been considered~\cite{Popov:2005uv,Adamo:2017xaf,Bittleston:2020hfv,Bittleston:2025gxr} with the underlying complex manifold $\mathsf{E} = \mathsf{Tot} \big( \cO(2) \to \CP^1 \big)$. This complex 2-manifold is also known as minitwistor space $\mathbb{MT} = \mathsf{Tot} \big( \cO(2) \to \CP^1 \big)$ and it appears as a 1d reduction of twistor space $\mathbb{PT} = \mathsf{Tot} \big( \cO(1) \oplus \cO(1) \to \CP^1 \big)$. Non-commutative 5d CS theory on other manifolds has also been considered in \cite{Costello:2016nkh} in the context of describing M5 branes in an $\Omega$-background. The degree to which the holomorphic-topological theory is sensitive to the choice of $\mathsf{E}$, as well as any implications for the associated lower-dimensional integrable field theories, has not been systematically studied.

For the moment, our discussion will not be sensitive to global features such as the choice of bundle over $\CP^1$, so instead we work locally where our 5-manifold looks like $\Complex^2 \times \Real$. From this perspective, there are two distinct choices of reduction: a reduction along the complex directions (we refer to this as a holomorphic reduction), or a reduction along the real direction (we refer to this as a topological reduction). If we describe this reduction by the flow along a real vector field $V$, then we can distinguish the two cases by $V \contract \omega_2 \neq 0$ (holomorphic) and $V \contract \omega_2 = 0$ (topological).

Let us start by reviewing the holomorphic reduction presented in~\cite{Bittleston:2020hfv} that takes us to 4d CS. This computation is made easier by using the shift symmetry $A \mapsto A + C$ (where $\omega_2 \wedge C = 0$) to impose the condition $V \contract A = 0$. Notably, this is only possible because the vector field satisfies $V \contract \omega_2 \neq 0$; if it were a topological reduction, then $V \contract \omega_2 = 0$ implies $V \contract C = 0$ and it would not be possible to impose $V \contract A = 0$ using the shift symmetry.
We continue the reduction by imposing the constraint $\LieD_V A = 0$, and then contracting $V$ into the 5-form Lagrangian to get a 4-form that will be the Lagrangian of our 4d theory. Due to the constraints, the only non-trivial contribution comes from the contraction of $V$ with the meromorphic $(2,0)$-form, and the 4d theory is given by
\begin{equation}
\Act_{\text{CS}_4} [A] = \frac{1}{4 \pi \rmi} \int \omega_1 \wedge \bl{A}{\dr A + \frac{1}{3} [A, A]} ~, \qquad
\omega_1 = V \contract \omega_2 ~.
\end{equation}
This relationship between 5d CS and 4d CS was first shown by Bittleston and Skinner \cite{Bittleston:2020hfv}.

By comparison, let us consider a topological reduction where the vector field satisfies $V \contract \omega_2 = 0$. In this case, it is helpful to define a dual 1-form $\dr \rho$ satisfying $V \contract \dr \rho = 1$, in other words the coordinate $\rho$ parametrises the integral curves of $V = \pd_\rho$. We can decompose the gauge field into the component that points along these integral curves, which we denote by $b$, and the remaining components, denoted by $A^\perp$,
\begin{equation}
A = A^\perp + b \, \dr \rho ~, \qquad
b = V \contract A ~.
\end{equation}
Substituting this into the 5d CS action gives
\begin{equation}
\Act_{\text{CS}_5} [A] = \Act_{\text{CS}_5} [A^\perp] + \frac{1}{2 \pi \rmi} \int \dr \rho \wedge \omega_2 \wedge \bl{b}{F(A^\perp)} + \frac{1}{4 \pi \rmi} \int \dr \rho \wedge \dr \omega_2 \wedge \bl{b}{A^\perp} ~.
\end{equation}
To complete the reduction, we impose the constraint $\LieD_V A = 0$ and contract $V$ into the 5-form Lagrangian. The only non-trivial contribution comes from the contraction of $V$ with the explicit factors of $\dr \rho$, and we land on the action of 4d hBF theory,
\begin{equation}
\Act_{\text{hBF}_4} [b, A^\perp] = \frac{1}{2\pi \rmi} \int \omega_2 \wedge \bl{b}{F(A^\perp)} + \frac{1}{4 \pi \rmi} \int \dr \omega_2 \wedge \bl{b}{A^\perp} ~,
\end{equation}
with a specific choice of boundary term.
Unlike the reduction to 4d CS theory, this reduction does not modify the choice of $\omega_2$, and hence the underlying complex manifold $\mathsf{E}$ will remain unchanged. Therefore, in order to recover the precise version of 4d hBF theory that we study in this paper, we would have to choose the underlying complex manifold $\mathsf{E} = \mathsf{Tot} \big( \cO(1) \to \CP^1 \big)$ in the initial 5d CS setup, which in turn can be found as a reduction of 6d hCS theory with underlying complex manifold $\mathsf{Tot}\big(\cO(0)\oplus\cO(1)\to\CP^1\big)$. This choice differs from the typical choice made in the literature~\cite{Popov:2005uv,Adamo:2017xaf,Bittleston:2020hfv}.

\section{Defects and integrable models}\label{sec:defects}

In this section we consider 4d hBF with defects and the associated integrable models.

\subsection{Simple pole setup}\label{sec:defects4dBF}

Let us consider an example setup in 4d hBF theory with the underlying complex manifold $\cO(1)$. For this example, we take the $(2,0)$-form to be
\begin{equation}\label{eq:omega2simplepolesetup}
\omega_2 = \frac{z - \gamma}{(z - \alpha)(z - \tilde{\alpha})} \, \dr z \wedge \dr w ~.
\end{equation}
This $(2,0)$-form\footnote{In homogeneous coordinates, this $(2,0)$-form is written as $\omega_2 = \frac{\langle \lambda \gamma \rangle}{\langle \lambda \alpha \rangle \langle \lambda \tilde{\alpha} \rangle \langle \lambda \beta \rangle} \langle \lambda \dr \lambda \rangle \wedge \lambda_a \dr x^a$ where $\langle \lambda \gamma \rangle = \lambda^a \gamma_a$.} has two finite simple poles at $z = \alpha$ and $z = \tilde{\alpha}$ and a simple pole at $z = \infty$. To see the pole at infinity, we should go to the other patch parametrised by $(\tilde{z}, \tilde{w})$ using the maps $\tilde{z} = 1/z$ and $\tilde{w} = w/z$ and take into account that $\omega_2$ itself transforms as $\omega_2 \mapsto \omega_2 / z$ because it is valued in $\cO(1)$. In addition to the poles, we must also pay attention to the zeroes of $\omega_2$, and this $(2,0)$-form has one finite zero at $z = \gamma$.

In order to complete our setup, we need to provide boundary conditions on the fundamental fields at the poles of $\omega_2$ and specify any allowed singularities in the field configurations at the zeroes of $\omega_2$. At a simple pole $z_p$ of $\omega_2$, the boundary variation of the action will include a boundary term of the form
\begin{equation}
\int_{\Real^2} \dr w \wedge \bl{\delta A}{b} \big\vert_{z = z_p} ~.
\end{equation}
We will ensure that this vanishes by imposing the boundary condition $A_{\hat{w}} \vert_{z = z_p} = 0$ at each simple pole of $\omega_2$. It is important to highlight that this boundary condition breaks the gauge invariance of 4d hBF theory at the poles of $\omega_2$ leading to the emergence of local degrees of freedom at these points. Turning to the zeroes of $\omega_2$, we will allow both $A$ and $b$ to have a simple pole at $z = \gamma$. Normally, allowing both of these fields to have a pole at the same point would overwhelm the zero in $\omega_2$, but we circumvent this by imposing the constraint
\begin{equation}\label{eq:res1}
\res_\gamma (b) = q_\gamma \cdot \res_\gamma (A_{\hat{w}}) ~.
\end{equation}
In this expression, $q_\gamma$ is an unspecified complex number measuring the ratio between the two residues and it will become a parameter in the 2d integrable field theory. This constraint ensures that the action remains finite despite both fields having a shared pole.

We can derive the associated 2d integrable theory by defining $A = \hat{g}^{-1} A^\prime \hat{g} + \hat{g}^{-1} \dr \hat{g}$ and $b = \hat{g}^{-1} b^\prime \hat{g}$. These field redefinitions introduce a redundancy into the theory acting as
\begin{equation}
A^\prime \mapsto \check{h}^{-1} A^\prime \check{h} + \check{h}^{-1} \dr \check{h} ~, \qquad
b^\prime \mapsto \check{h}^{-1} b^\prime \check{h} ~, \qquad
\hat{g} \mapsto \check{h}^{-1} \hat{g} ~.
\end{equation}
We will use this redundancy to impose the constraint $A^\prime_{\bar{z}} = 0$ such that two of the equations of motion become $\varphi \cdot \pd_{\bar{z}} A^\prime_{\hat{w}} = 0$ and $\varphi \cdot \pd_{\bar{z}} b^\prime = 0$. These equations imply that $A^\prime_{\hat{w}}$ and $b^\prime$ are holomorphic everywhere except for $z = \gamma$ where we have allowed them to have a pole. The most general solutions to these equations satisfying the constraint~\eqref{eq:res1} can be written\footnote{In homogeneous coordinates, these solutions are $A^\prime_{\hat{w}} = \langle \lambda \hat{\beta} \rangle \, a_\infty + \frac{\langle \lambda \beta \rangle \langle \lambda \tilde{\alpha} \rangle}{\langle \lambda \gamma \rangle} a_\alpha + \frac{\langle \lambda \beta \rangle \langle \lambda \alpha \rangle}{\langle \lambda \gamma \rangle} a_{\tilde{\alpha}}$ and $b^\prime = q_\gamma \frac{\res_\gamma (A^\prime_{\hat{w}})}{\langle \lambda \gamma \rangle}$.} as
\begin{equation}
A^\prime_{\hat{w}} = z \, a_\infty + \frac{z - \tilde{\alpha}}{z - \gamma} a_\alpha + \frac{z - \alpha}{z - \gamma} a_{\tilde{\alpha}} ~, \qquad
b^\prime = q_\gamma \frac{\res_\gamma (A^\prime_{\hat{w}})}{z - \gamma} ~.
\end{equation}
When writing these expressions, we have also taken into account the global behaviour of the fields. Since the basis $1$-form $e^{\hat{w}}$ accompanying $A^\prime_{\hat{w}}$ goes like $z^{-1}$ at infinity, the gauge field remains finite as long as this component is at most linear in $z$. Meanwhile, $b^\prime$ is valued in $\cO(-1)$, which has no global holomorphic sections, so only the singular piece survives.

We would like to solve for the three unknowns $a_\infty$, $a_\alpha$ and $a_{\tilde{\alpha}}$ in terms of the edge mode field $\hat{g}$. Before doing this, it is helpful to denote the relevant degrees of freedom by
\begin{equation}
\hat{g} \vert_\alpha = g ~, \qquad
\hat{g} \vert_{\tilde{\alpha}} = \tilde{g} ~, \qquad
\hat{g} \vert_\infty = \id ~.
\end{equation}
At infinity, we have used the residual transformations parametrised by $\check{h}$ that preserve the constraint $A^\prime_{\bar{z}} = 0$ to gauge fix this degree of freedom. It is also helpful to introduce some more notation: let us denote by $\pd_\alpha$ the vector field $\pd_{\hat{w}} = \pd_2 - z \, \pd_1$ evaluated at $z = \alpha$ and similarly for the other poles with the limit $z \to \infty$ denoted by $\pd_\beta = -\pd_1$. We can now solve the boundary conditions for the unknowns to find the on-shell configurations
\begin{equation}\label{eq:Ab}\begin{aligned}
A^\prime_{\hat{w}} & = \frac{(z - \tilde{\alpha})(\alpha - \gamma)(-\pd_\alpha g g^{-1}) - (z - \alpha)(\tilde{\alpha} - \gamma)(-\pd_{\tilde{\alpha}} \tilde{g} \tilde{g}^{-1})}{(z - \gamma)(\alpha - \tilde{\alpha})} ~, \\
b^\prime & = q_\gamma \frac{(\alpha - \gamma)(\tilde{\alpha} - \gamma)}{\alpha - \tilde{\alpha}} \frac{\pd_\alpha g g^{-1} - \pd_{\tilde{\alpha}} \tilde{g} \tilde{g}^{-1}}{z - \gamma} ~.
\end{aligned}\end{equation}
We will substitute these field configurations back into the action to derive the 2d integrable theory.

In the gauge we are working in, the action takes the form
\begin{equation}
\Act_{\text{hBF}_4} = \frac{1}{2 \pi \rmi} \int \omega_2 \wedge \bl{b^\prime}{\pdb_{\CP^1} A^\prime} ~.
\end{equation}
Since $A^\prime$ is meromorphic with a pole at $z = \gamma$, the term $\pdb_{\CP^1} A^\prime$ is a distribution with support at this point. One might think that this contribution vanishes due to the zero in $\omega_2$, but this is counteracted by the coincident pole in $b^\prime$. This means that the action becomes
\begin{equation}\begin{split}\label{eq:2dift}
\Act_{\mathrm{2dIFT}}[g,\tilde g] & =
\frac{q_\gamma}{(\alpha - \gamma)(\tilde{\alpha} - \gamma)} \int_{\Real^2} \dr^2 x \, \bl{\res_\gamma (A^\prime_{\hat{w}})}{\res_\gamma (A^\prime_{\hat{w}})} \\ & =
\frac{q_\gamma(\alpha - \gamma) (\tilde{\alpha} - \gamma)}{(\alpha - \tilde{\alpha})^2} \int_{\Real^2} \dr^2 x \, \bl{\pd_\alpha g g^{-1} - \pd_{\tilde{\alpha}} \tilde{g} \tilde{g}^{-1}}{\pd_\alpha g g^{-1} - \pd_{\tilde{\alpha}} \tilde{g} \tilde{g}^{-1}} ~,
\end{split}\end{equation}
where $ \dr^2 x = \dr x^1 \wedge \dr x^2$.
This is the action of the 2d integrable field theory and its equations of motion can be found by varying $g$ and $\tilde{g}$ independently. They are equivalent to the equation $\pd_{\hat{w}} b^\prime + [A^\prime_{\hat{w}}, b^\prime] = 0$ holding for all values of $z \in \CP^1$.

This 2d theory is invariant under three chiral symmetries acting as
\begin{equation}\begin{gathered}
g \mapsto h_\beta^{-1} \cdot g \cdot h_\alpha ~, \qquad
\tilde{g} \mapsto h_\beta^{-1} \cdot \tilde{g} \cdot h_{\tilde\alpha} ~, \\
\pd_\alpha h_\alpha = 0 ~, \qquad
\pd_{\tilde\alpha} h_{\tilde\alpha} = 0 ~, \qquad
\pd_\beta h_\beta = 0 ~.
\end{gathered}\end{equation}
These can be identified with the 4d hBF theory gauge transformations that preserve the boundary conditions $A \vert_{z = z_p} = 0$. The currents associated with these symmetries are given by
\begin{equation}\begin{aligned}
\pd_\alpha J_\alpha & = 0 ~, & \qquad
J_\alpha & = g^{-1} \big( \pd_\alpha g g^{-1} - \pd_{\tilde{\alpha}} \tilde{g} \tilde{g}^{-1} \big) g ~, \\
\pd_{\tilde\alpha} J_{\tilde\alpha} & = 0 ~, & \qquad
J_{\tilde\alpha} & = \tilde{g}^{-1} \big( \pd_\alpha g g^{-1} - \pd_{\tilde{\alpha}} \tilde{g} \tilde{g}^{-1} \big) \tilde{g} ~, \\
\pd_\beta J_\beta & = 0 ~, & \qquad
J_\beta & = \pd_\alpha g g^{-1} - \pd_{\tilde{\alpha}} \tilde{g} \tilde{g}^{-1} ~.
\end{aligned}\end{equation}
These currents can be identified with the evaluation of $b$ at each pole of $\omega_2$.

\subsection{Analysis of the 2d IFT}\label{sec:analysis}

In the following sections, we will show that this 2d holomorphic IFT can be recovered as the limit of a typical 2d IFT (constructed from 4d CS) and reduces to a typical 1d IM (constructed from 3d BF).
Before exploring these comparisons, let us perform an initial classical analysis of the 2d holomorphic IFT~\eqref{eq:2dift}.
Varying the action with respect to $g$ and $\tilde{g}$ gives equations of motion that are equivalent to the conservation laws of the chiral currents $J_\alpha = g^{-1} J_\beta g$ and $J_{\tilde \alpha} = \tilde{g}^{-1} J_\beta \tilde{g}$,
\begin{equation}
\pd_\alpha \big( g^{-1} J_\beta g \big) = 0 ~, \qquad
\pd_{\tilde\alpha} \big( \tilde{g}^{-1} J_\beta \tilde{g} \big) = 0 ~, \qquad
J_\beta = \pd_\alpha g g^{-1} - \pd_{\tilde{\alpha}} \tilde{g} \tilde{g}^{-1} ~.
\end{equation}
The third conservation law
\begin{equation}
\partial_\beta J_\beta = 0 ~,
\end{equation}
is implied by the other two equations.
Let us study these equations (in the case that $\alg{g}$ is the compact real form of a complex semisimple Lie algebra) to understand the space of solutions of the 2d theory.
It will turn out that we can exactly solve these equations and write down an infinite family of solutions, consistent with the classical integrability of this model.

Consider $\mathsf{P} (J_\beta)$ where $\mathsf{P} : \alg{g} \to \Complex$ is an arbitrary ad-invariant polynomial on the Lie algebra.
Since each of the chiral currents is related by an adjoint action, their conservation laws imply that $\mathsf{P} (J_\beta)$ is constant in spacetime.
This is a consequence of the holomorphic integrability of the 2d IFT and $\omega_2$~\eqref{eq:omega2simplepolesetup} having only one zero.
The residue of $b^\prime$ at $z = \gamma$ is proportional to $J_\beta$, so we can factor out this $z$-dependence and rewrite the integrability condition as
\begin{equation}
\pd_{\hat{w}} J_\beta + [A^\prime_{\hat{w}}, J_\beta] = 0 ~.
\end{equation}
This equation implies that $\pd_{\hat{w}} \mathsf{P} (J_\beta) = 0$ for all values of $z$, which further implies that $\mathsf{P} (J_\beta)$ must be constant since $\pd_{\hat{w}} = \pd_2 - z \, \pd_1$ spans the entire tangent space of spacetime as $z$ varies over $\CP^1$.
In effect, $\mathsf{P} (J_\beta)$ are conserved quantities associated with the integrability of this theory.
For setups with more than one zero in $\omega_2$ this argument will not hold and the conserved quantities will be integrals over space, akin to standard 2d IFTs.
We will comment further on this in the concluding remarks.

The fact that $\mathsf{P} (J_\beta)$ is constant in spacetime is a significant restriction on the space of solutions: each conjugacy class of the Lie algebra $\alg{g}$ is uniquely determined by these invariants, so $J_\beta$ must live in a fixed conjugacy class across all of spacetime.
Every element of a fixed conjugacy class can be written as $J_\beta = h^{-1} \cdot \Gamma_0 \cdot h$ where $\Gamma_0 \in \alg{t}$ is a fixed element of the Cartan subalgebra $\alg{t} = \mathrm{CSA}(\alg{g})$.
Here the map $h : \Real^2 \to \grp{G}$ can have arbitrary dependence on spacetime as the operators $\mathsf{P} (J_\beta)$ are insensitive to this choice -- they only measure properties of $\Gamma_0 \in \alg{t}$.
We can restrict this spacetime dependence by returning to the chiral conservation laws.

The first conservation law $\pd_\beta J_\beta = 0$ implies that the map $h$ takes the form $h = f_0 \cdot h_\beta$ where the second element is chiral $\pd_\beta h_\beta = 0$, and the first element has arbitrary spacetime dependence but always lies in the stabiliser $\grp{G}_{\Gamma_0}$ of $\Gamma_0$, meaning $f_0^{-1} \cdot \Gamma_0 \cdot f_0 = \Gamma_0$. For a generic (regular) element of the Cartan subalgebra, this stabiliser is the maximal torus $\grp{T} \subset \grp{G}$ found by exponentiating the Cartan subalgebra $\grp{T} = \exp (\alg{t})$.

Turning to the remaining conservation laws, they imply that the combinations $h \cdot g$ and $h \cdot \tilde{g}$ can be written in a similar fashion, which in turn determines explicit expressions for the fundamental fields,
\begin{equation}\begin{aligned}
\pd_\alpha \big( g^{-1} J_\beta g \big) & = 0 \qquad \implies \qquad
h \cdot g = f_0^\prime \cdot h_\alpha \qquad \implies \qquad
g = h_\beta^{-1} \cdot f_0^{-1} \cdot f_0^\prime \cdot h_\alpha ~, \\
\pd_{\tilde\alpha} \big( \tilde{g}^{-1} J_\beta \tilde{g} \big) & = 0 \qquad \implies \qquad
h \cdot \tilde{g} = f_0^{\prime \prime} \cdot h_{\tilde\alpha} \qquad \implies \qquad
\tilde{g} = h_\beta^{-1} \cdot f_0^{-1} \cdot f_0^{\prime \prime} \cdot h_{\tilde\alpha} ~.
\end{aligned}\end{equation}
In these expressions, the maps $h_\alpha : \Real^2 \to \grp{G}$ and $h_{\tilde\alpha} : \Real^2 \to \grp{G}$ are required to satisfy the chirality constraints $\pd_\alpha h_\alpha = 0$ and $\pd_{\tilde\alpha} h_{\tilde\alpha} = 0$, and $(f_0^\prime , f_0^{\prime \prime})$ are required to lie in the stabiliser $\grp{G}_{\Gamma_0}$ of $\Gamma_0$. Since the fundamental fields only depend on the various elements of the stabiliser through specific combinations, we can redefine $f_0$, $f_0'$ and $f_0''$ such that
\begin{equation}\label{eq:factorisation}
g = h_\beta^{-1} \cdot f_0 \cdot h_\alpha ~, \qquad
\tilde{g} = h_\beta^{-1} \cdot \tilde{f}_0 \cdot h_{\tilde\alpha} ~.
\end{equation}
Subject to $J_\beta = h_\beta^{-1}\Gamma_0h_\beta$, these constitute the generic on-shell field configurations for the fundamental fields of the 2d theory.

In summary, the components $(h_\beta, h_\alpha, h_{\tilde\alpha}, f_0 , \tilde{f}_0)$ are all maps from spacetime to the group, and they satisfy the constraints
\begin{equation}
\pd_\beta h_\beta = 0 ~, \qquad
\pd_\alpha h_\alpha = 0 ~, \qquad
\pd_{\tilde\alpha} h_{\tilde\alpha} = 0 ~, \qquad
f_0 (x) \in \grp{G}_{\Gamma_0} \subset \grp{G} ~, \qquad
\tilde{f}_0 (x) \in \grp{G}_{\Gamma_0} \subset \grp{G} ~.
\end{equation}
For these field configurations, the constant combination $\Gamma_0$ is given by
\begin{equation}
\Gamma_0 = \pd_\alpha f_0^{\phantom{1}} f_0^{-1} - \pd_{\tilde{\alpha}} \tilde{f}_0^{\phantom{1}} \tilde{f}_0^{-1} ~, \qquad
\dr \Gamma_0 = 0 ~.
\end{equation}
The condition $\dr \Gamma_0 = 0$ imposes an additional constraint on the functions $(f_0 , \tilde{f}_0)$ in the on-shell field configurations.
Before imposing this constraint, there are two free functions in the on-shell configurations, so we are left with one free function after demanding that $\Gamma_0$ is constant.
This feature is somewhat unusual, so let us take a moment to investigate it.

Consider a generic $\Gamma_0$ such that the stabiliser $\grp{G}_{\Gamma_0} = \grp{T}$, the Cartan torus.
We will restrict our attention to the abelian subsector of the on-shell field configurations parametrised by $f_0$ and $\tilde{f}_0$.
Making use of the chiral symmetries in this theory, it is possible to transform any solution into a field configuration which lies exclusively in the maximal torus $\grp{T} \subset \grp{G}$.
Conversely, we can construct the entire space of solutions to the theory by considering maps from spacetime into the maximal torus and then acting on these field configurations with the chiral symmetries.
With this in mind, we will set the chiral components to the identity, and evaluate the action on the configurations $f_0 = \exp (Y)$ and $\tilde{f}_0 = \exp (\tilde{Y})$ where $Y(x) \in \alg{t}$ and $\tilde{Y}(x) \in \alg{t}$. This gives
\begin{equation}
\Act_{\text{IFT}_2}[Y,\tilde Y] = \frac{q_\gamma(\alpha - \gamma) (\tilde{\alpha} - \gamma)}{(\alpha - \tilde{\alpha})^2} \int_{\Real^2} \dr^2 x \, \bl{\pd_\alpha Y - \pd_{\tilde{\alpha}} \tilde{Y}}{\pd_\alpha Y - \pd_{\tilde{\alpha}} \tilde{Y}} ~.
\end{equation}
In this action there is a gauge symmetry acting on the fields as
\begin{equation}
Y \mapsto Y + \pd_{\tilde{\alpha}} \Theta ~, \qquad
\tilde{Y} \mapsto \tilde{Y} + \pd_\alpha \Theta ~, \qquad
\Theta : \Real^2 \to \alg{t} ~.
\end{equation}
This is the origin of the free function above, which is a symmetry of the on-shell configurations because it leaves both the (restricted) action and the combination $\Gamma_0 = \pd_\alpha Y - \pd_{\tilde{\alpha}} \tilde{Y}$ invariant.
It is unclear if this symmetry can be realised in the original variables $(g, \tilde{g})$ without first solving the equations of motion, and therefore it is unclear how to interpret this symmetry in the full nonabelian theory.
Moreover, if it were possible to realise this symmetry in the full nonabelian theory, since the gauge parameter is valued in the Cartan subalgebra, it could not be used to set $\tilde g = 1$, and it would not render the theory a complexification of a 1d IM as it does in the abelian case.

We can also observe the presence of this gauge symmetry in the Hamiltonian formalism.
Let us define our spacetime coordinates such that $(x^1, x^2) = (x, t)$ and compute the conjugate momenta for the two fields.
Setting
\begin{equation}
q_\gamma = \frac{(\alpha-\tilde\alpha)^2}{(\alpha-\gamma)(\tilde\alpha-\gamma)} ~,
\end{equation}
the conjugate momenta of $g$ and $\tilde{g}$ are given by
\begin{equation}
\pi = J_\beta ~, \qquad
\tilde{\pi} = -J_\beta ~, \qquad
J_\beta = \pd_t g g^{-1} - \alpha \, \pd_x g g^{-1} - \pd_t \tilde{g} \tilde{g}^{-1} + \tilde\alpha \, \pd_x \tilde{g} \tilde{g}^{-1} ~.
\end{equation}
These relations imply that $\pi + \tilde{\pi} \approx 0$ is a primary constraint on our system. The Hamiltonian density is given by
\begin{equation}
\Ham = \frac{1}{2} \bl{J_\beta}{J_\beta} + \bl{J_\beta}{\alpha \, \pd_x g g^{-1} - \tilde\alpha \, \pd_x \tilde{g} \tilde{g}^{-1}} ~.
\end{equation}
Due to the constraint, there is no unique way to rewrite this in terms of $\pi$ and $\tilde{\pi}$. Let us arbitrarily choose to always replace $J_\beta$ with $\pi$ in the Hamiltonian.

In tensor notation, where $C_\tn{12}$ is the quadratic Casimir, the fundamental Poisson brackets are given by
\begin{equation}\begin{aligned}
\{ \pi(x)_\tn{1} , g(y)_\tn{2} \} & = -C_\tn{12} \cdot g(y)_\tn{2} \, \delta(x-y) ~, & \qquad &&
\{ \pi(x)_\tn{1} , \pi(y)_\tn{2} \} & = [C_\tn{12} , \pi(x)_\tn{1}] \, \delta(x-y) ~, \\
\{ \tilde{\pi}(x)_\tn{1} , \tilde{g}(y)_\tn{2} \} & = -C_\tn{12} \cdot \tilde{g}(y)_\tn{2} \, \delta(x-y) ~, & \qquad &&
\{ \tilde{\pi}(x)_\tn{1} , \tilde{\pi}(y)_\tn{2} \} & = [C_\tn{12} , \tilde{\pi}(x)_\tn{1}] \, \delta(x-y) ~.
\end{aligned}\end{equation}
We can use these to compute the Poisson bracket of the momenta with the spatial components of the Maurer-Cartan forms,
\begin{equation}\begin{aligned}
\{ \pi(x)_\tn{1} , \pd_y g g^{-1} (y)_\tn{2} \} & = [C_\tn{12} , \pd_y g g^{-1} (y)_\tn{1} ] \delta(x-y) + C_\tn{12} \, \delta^\prime (x-y) ~, \\
\{ \tilde{\pi}(x)_\tn{1} , \pd_y \tilde{g} \tilde{g}^{-1} (y)_\tn{2} \} & = [C_\tn{12} , \pd_y \tilde{g} \tilde{g}^{-1} (y)_\tn{1} ] \delta(x-y) + C_\tn{12} \, \delta^\prime (x-y) ~.
\end{aligned}\end{equation}
Computing the Poisson bracket of the Hamiltonian with the primary constraint $\pi + \tilde{\pi} \approx 0$ to see if it generates any secondary constraints, we find
\begin{equation}
\int \dr y \, \{ \Ham(y) , \pi(x) + \tilde{\pi}(x) \} = (\tilde\alpha - \alpha) \, \pd_x \pi(x) ~.
\end{equation}
Therefore, we have the secondary constraint $\pd_x \pi \approx 0$.
Proceeding, we find the tertiary constraint $\pd_x^2 \pi \approx 0$, but this is already implied by the secondary constraint.
At this point we can consider the Poisson bracket between the two constraints
\begin{equation}
\{\pi(x)_\tn{1}+\tilde\pi(x)_\tn{1}, \pd_y \pi(y)_\tn{2}\} = [C_\tn{12},\pi(x)_\tn{1}]\partial_y\delta(x-y) ~.
\end{equation}
Since the adjoint action on the Lie algebra has a kernel, this Poisson bracket is not invertible indicating the presence of gauge symmetries.
Moreover, the kernel is the stabiliser of $\pi(x) = J_\beta$ mirroring the Lagrangian analysis.

Understanding this gauge symmetry and how to treat it is important for quantizing the theory. Indeed, a naive perturbative analysis, i.e., parametrising $g=\exp(\Phi)$ and $\tilde g = \exp(\tilde \Phi)$, and expanding leads to a non-invertible quadratic term, again indicating the presence of gauge symmetries.
We will comment on this further in the concluding remarks.

Let us conclude this Hamiltonian analysis by noting that if we compute the Poisson bracket of $b'$~\eqref{eq:Ab} with itself at equal $t$ using the above results, we find an ultralocal Poisson bracket similar to the general analysis in \secref{sec:hamiltonian4dhBF}.
However, we do not simply recover~\eqref{eq:ultralocal} with $w$ and $\bar w$ replaced by $x$ and $t$ respectively.
In particular, we are left with a different dependence on $z_1$ and $z_2$ on the right-hand side.
This might be expected since in this example we are working with $\cO(1) \to \CP^1$ and the map between $(w,\bar w)$ and the real coordinates of 2d spacetime now depends on the $\CP^1$ coordinate $z$.
While we leave a detailed study for the future, this suggests two approaches to continuing the analysis of the classical integrability of holomorphic 2d IFTs.
The first is to compute the Poisson bracket of $b^\prime$ with itself at equal times.
The second is to take~\eqref{eq:ultralocal} at equal $\hat{w}$ as a novel Poisson bracket and study the implications for the algebra of conserved charges.

\subsection{As a limit of 4d CS}\label{sec:defects4dCS}

Let us consider a 4d CS setup that reduces to the 4d hBF example setup from~\secref{sec:defects4dBF}. Before directly constructing this 4d CS setup, we will show that it can be understood as a generalisation of the more typical relativistic setups commonly considered in the literature. Since this relativistic setup is not the main focus of our analysis, we will be brief when presenting this model.

We define 4d CS theory over the manifold $\CP^1 \times \Real^2$ with coordinates $z \in \CP^1$ and $x^a \in \Real^2$ for $a \in \{1, 2\}$. The action of 4d CS is given by
\begin{equation}
\Act_{\text{CS}_4} [A] = \frac{k}{4 \pi \rmi} \int \omega_1 \wedge \bl{A}{\dr A + \frac{1}{3} [A, A]} ~.
\end{equation}
For the relativistic setup, we take the meromorphic $(1,0)$-form to be
\begin{equation}
\omega_1 = \frac{(z - \gamma)(z - \tilde{\gamma})}{(z - \alpha)(z - \tilde{\alpha})} \, \dr z ~.
\end{equation}
This has two finite simple poles at $z = \alpha$ and $z = \tilde\alpha$ and a double pole at $z = \infty$. It also has two simple zeroes at $z = \gamma$ and $z = \tilde\gamma$. To specify the boundary conditions and singularities of $A$ at these points, we momentarily swap to lightcone coordinates $x^\pm = x^1 \pm x^2$. We will allow $A_+$ to have a pole at $z = \gamma$ and the other spacetime component $A_-$ to have a pole at $z = \tilde\gamma$. We will then impose boundary conditions on the gauge field given by
\begin{equation}
A_+ \vert_{z = \alpha} = 0 ~, \qquad
A_- \vert_{z = \tilde\alpha} = 0 ~, \qquad
A_\pm \vert_{z = \infty} = 0 ~.
\end{equation}
By following a similar analysis to the other localisation calculations in this paper, the action of the associated 2d integrable theory is given by
\begin{equation}\begin{aligned}
\Act_{\text{IFT}_2} [g, \tilde{g}] & =
k \, \frac{(\alpha - \gamma)(\alpha - \tilde\gamma)}{\alpha - \tilde\alpha} \, \bigg[ 2\int \dr^2 x \,\bl{g^{-1} \pd_+ g}{g^{-1} \pd_- g} - \frac{1}{6} \int \bl{g^{-1} \dr g}{[g^{-1} \dr g, g^{-1} \dr g]} \bigg] \\
& \quad + k \, \frac{(\tilde\alpha - \gamma)(\tilde\alpha - \tilde\gamma)}{\alpha - \tilde\alpha} \, \bigg[ 2 \int \dr^2 x \, \bl{\tilde{g}^{-1} \pd_+ \tilde{g}}{\tilde{g}^{-1} \pd_- \tilde{g}} + \frac{1}{6} \int \bl{\tilde{g}^{-1} \dr \tilde{g}}{[\tilde{g}^{-1} \dr \tilde{g}, \tilde{g}^{-1} \dr \tilde{g}]} \bigg] \\
& \quad -4 k \, \frac{(\alpha - \gamma)(\tilde\alpha - \tilde\gamma)}{\alpha - \tilde\alpha} \, \int \dr^2 x \, \bl{\pd_+ g g^{-1}}{\pd_- \tilde{g} \tilde{g}^{-1}} ~,
\end{aligned}\end{equation}
where $\dr^2 x = \dr x^1 \wedge \dr x^2 = -\frac{1}{2} \dr x^+ \wedge \dr x^-$.
This theory describes two Wess-Zumino-Witten (WZW) models coupled by a relativistic quadratic term. Neglecting for one moment the quadratic coupling, the two WZW models are invariant under chiral symmetries acting as
\begin{equation}\begin{gathered}
g \mapsto h_\ell^{-1} \cdot g \cdot h_r ~, \qquad
\pd_- h_\ell = 0 ~, \qquad \pd_+ h_r = 0 ~, \\
\tilde{g} \mapsto \tilde{h}_\ell^{-1} \cdot \tilde{g} \cdot \tilde{h}_r ~, \qquad
\pd_+ \tilde{h}_\ell = 0 ~, \qquad \pd_- \tilde{h}_r = 0 ~.
\end{gathered}\end{equation}
The left and right-acting symmetries are constant with respect to different directions in these two models due to the different relative sign between the kinetic term and the Wess-Zumino (WZ) term. In order to retain this information but also write the action more concisely, we will introduce some notation. We write the action of the 2d integrable theory as
\begin{equation}
\Act_{\text{IFT}_2} [g, \tilde{g}] =
k \, r_\alpha \, \Act_{\text{WZW}_{+-}} [g]
+ k \, r_{\tilde\alpha} \, \Act_{\text{WZW}_{-+}} [\tilde{g}]
- 4 k \, \frac{(\alpha - \gamma)(\tilde\alpha - \tilde\gamma)}{\alpha - \tilde\alpha} \, \int \dr^2 x \, \bl{\pd_+ g g^{-1}}{\pd_- \tilde{g} \tilde{g}^{-1}} ~.
\end{equation}
The subscripts on the WZW models indicate the constraints on the right-acting and left-acting symmetries, and the relative overall sign is determined by $\Act_{\text{WZW}_{+-}} [g] = - \Act_{\text{WZW}_{-+}} [g^{-1}]$.
We have also denoted the residues of $\omega_1$ at the two simple poles by $r_\alpha$ and $r_{\tilde\alpha}$ respectively. Taking into account the coupling term, the full theory is invariant under transformations $g \mapsto g \cdot h_\alpha$ where $\pd_+ h_\alpha = 0$ and transformations $\tilde{g} \mapsto \tilde{g} \cdot h_{\tilde\alpha}$ where $\pd_- h_{\tilde\alpha} = 0$. These symmetries can be understood as the gauge transformations of 4d CS theory that preserve the boundary conditions. The inclusion of the coupling term preserves half of the chiral symmetries but breaks the other half.

We would like to generalise this setup by breaking Lorentz invariance and writing everything in terms of $A_w$ and $A_{\hat{w}}$. This will prepare the setup for taking the limit to 4d hBF theory. First, let us take the zero at $z = \tilde\gamma$ and send it to infinity $\tilde\gamma \to \infty$. This will turn the double pole at infinity into a simple pole and leave us with the meromorphic $(1,0)$-form
\begin{equation}
\omega_1 = \frac{z - \gamma}{(z - \alpha)(z - \tilde{\alpha})} \, \dr z ~.
\end{equation}
To write boundary conditions that are adapted to the 4d hBF limit, we introduce the coordinates $w = x^1 + z \, x^2$ and $\hat{w} = (x^2 - \bar{z} \, x^1)/(1 + z \bar{z})$. It should be emphasised that we are still working with standard 4d CS theory with underlying complex manifold $\CP^1$, and the $\cO(1)$-inspired coordinates are simply notation. We write the gauge field as
\begin{equation}
A = A_{\bar{z}} \dr \bar{z} + A_{\hat{w}} e^{\hat{w}} + A_w e^w ~, \qquad
e^{\hat{w}} = \frac{\dr x^2 - \bar{z} \, \dr x^1}{1 + z \bar{z}} ~, \qquad
e^w = \dr x^1 + z \, \dr x^2 ~.
\end{equation}
In this notation, we expect the components $A_{\bar{z}} \dr \bar{z} + A_{\hat{w}} e^{\hat{w}}$ to become the 4d hBF gauge field, while the component $A_w$ will become $\frac{1}{k} b$, which diverges as $k \to 0$. For this reason, we will only impose boundary conditions on $A_{\hat{w}}$ so that they remain finite in the limit. At each simple pole $z_p$ of $\omega_1$, we will impose the boundary condition $A_{\hat{w}} \vert_{z_p} = 0$, matching the 4d hBF theory boundary conditions from the example above in the limit $k \to 0$.

When it comes to the zero of $\omega_1$ at $z = \gamma$, we would like to allow some of the components of $A$ to have singularities at this point. It is easier to understand the limit of these singularities if we phrase the choice as a constraint, so we start by allowing both spacetime components of $A$ to have a singularity at $z = \gamma$. We then demand that some component of $A$ is regular at this point, ensuring that these singularities do not overwhelm the zero of $\omega_1$ in the action. Based on our previous discussion, we might anticipate that the correct component to constrain is $A_{\hat{w}}$ since this component remains finite in the limit to 4d hBF theory. However, if we impose all of the boundary conditions on this component and also deny it any singularities, we find that the system is overdetermined and we cannot solve for the on-shell configuration in terms of the edge modes. Said differently, we find that the boundary conditions have no solutions unless additional constraints are imposed on the edge modes.

The other way to get a finite object in the limit $k \to 0$ is to explicitly include $k$ in our constraint. Let us demand that the component $k \, A_{w} - q_\gamma \, A_{\hat{w}}$ is regular at $\gamma$, meaning
\begin{equation}\label{eq:res2}
\res_\gamma (k \, A_{w}) = q_\gamma \cdot \res_\gamma (A_{\hat{w}}) ~.
\end{equation}
Due to the explicit factor of $k$ in this constraint, it is well-behaved in the limit $k \to 0$ and reduces to the constraint we imposed in the 4d hBF setup. From the perspective of 4d CS theory, there is no need for this factor of $k$ to be related to the overall coefficient in front of the action, and this setup would be well-defined without this identification. If we were to take the limit $k \to 0$ in this residue condition, without also taking the overall coefficient in front of the action to zero, we would be approaching the overdetermined point in 4d CS theory, where too many constraints have been imposed on the component $A_{\hat{w}}$. By comparison, if we identify the two parameters and approach this overdetermined point while simultaneously taking the overall coefficient to zero, we recover the 4d hBF theory setup.

Let us derive the 2d integrable theory associated with this 4d CS setup. We define $A = \hat{g}^{-1} A^\prime \hat{g} + \hat{g}^{-1} \dr \hat{g}$ and fix the redundancy this introduces by imposing the constraints $A^\prime_{\bar{z}} = 0$ and $\hat{g} \vert_{\infty} = \id$. Solving the equations of motion in the basis $\{ e^w, e^{\hat{w}} \}$ requires care since these $1$-forms are not closed. The relevant equations are given by
\begin{equation}
\pd_{\bar{z}} A^\prime_{\hat{w}} = 0 ~, \qquad
\pd_{\bar{z}} A^\prime_w = \frac{A^\prime_{\hat{w}}}{(1 + z \bar{z})^2} ~.
\end{equation}
The most general solutions to these equations, which also satisfy the residue condition, are given by
\begin{equation}\begin{aligned}
A^\prime_{\hat{w}} & = z \, a_\infty + \frac{z - \tilde{\alpha}}{z - \gamma} a_\alpha + \frac{z - \alpha}{z - \gamma} a_{\tilde{\alpha}} ~, \\
A^\prime_w & = \frac{-1}{1 + z \bar{z}} \, a_\infty + \frac{\bar{z}}{1 + z \bar{z}} \bigg( \frac{z - \tilde{\alpha}}{z - \gamma} a_\alpha + \frac{z - \alpha}{z - \gamma} a_{\tilde{\alpha}} \bigg) + \bigg( \frac{q_\gamma}{k} - \frac{\bar{\gamma}}{1 + \gamma \bar{\gamma}} \bigg) \frac{\res_\gamma (A^\prime_{\hat{w}})}{z - \gamma} ~.
\end{aligned}\end{equation}
The field configuration for $A^\prime_{\hat{w}}$ identically matches that of 4d hBF theory, and solving the boundary conditions leads to the same expressions for the unknown components. On the other hand, the field configuration for $A^\prime_w$ has two distinct pieces: one that diverges in the limit $k \to 0$ due to the explicit factor of $1/k$, which is identified with the 4d hBF field $b$; and one that remains finite and is subleading in the limit.

Substituting the field redefinition into the 4d CS action gives
\begin{equation}
\Act_{\text{CS}_4} = \frac{k}{4 \pi \rmi} \int \omega_1 \wedge \bl{A^\prime}{\pdb_{\CP^1} A^\prime}
+ \frac{k}{4 \pi \rmi} \int \dr \omega_1 \wedge \bigg[ \bl{A^\prime}{\dr \hat{g} \hat{g}^{-1}} - \frac{1}{6} \int_{[0,1]} \bl{\hat{g}^{-1} \dr \hat{g}}{[\hat{g}^{-1} \dr \hat{g}, \hat{g}^{-1} \dr \hat{g}]} \bigg] ~.
\end{equation}
Unlike in the 4d hBF setup, the bulk term does not contribute to the final action once we substitute in the on-shell field configurations. The remaining terms are localised at the poles of $\omega_1$ and lead to the 2d action
\begin{equation}\begin{aligned}\label{eq:4dcssimple}
\Act_{\text{IFT}_2} [g, \tilde{g}] & = k \, r_\alpha \, \Act_{\text{WZW}_{\alpha \beta}} [g] + k \, r_{\tilde\alpha} \, \Act_{\text{WZW}_{\tilde\alpha \beta}} [\tilde{g}] \\
& + \frac{(\alpha - \gamma) (\tilde{\alpha} - \gamma)}{(\alpha - \tilde{\alpha})^2} \bigg( q_\gamma - \frac{k \, \bar{\gamma}}{1 + \gamma \bar{\gamma}} \bigg) \int \dr^2 x \, \bl{\pd_\alpha g g^{-1} - \pd_{\tilde{\alpha}} \tilde{g} \tilde{g}^{-1}}{\pd_\alpha g g^{-1} - \pd_{\tilde{\alpha}} \tilde{g} \tilde{g}^{-1}} ~.
\end{aligned}\end{equation}
While it might be surprising that the action appears to depend on $\bar\gamma$, this dependence can be removed by shifting $q_\gamma$ to remove $\bar\gamma$ from the residue condition written in terms of $A_1$ and $A_2$.
This theory is invariant under the symmetries
\begin{equation}\begin{gathered}
g \mapsto h_\beta^{-1} \cdot g \cdot h_\alpha ~, \qquad
\tilde{g} \mapsto h_\beta^{-1} \cdot \tilde{g} \cdot h_{\tilde\alpha} ~, \\
\pd_\alpha h_\alpha = 0 ~, \qquad
\pd_{\tilde\alpha} h_{\tilde\alpha} = 0 ~, \qquad
\pd_\beta h_\beta = 0 ~.
\end{gathered}\end{equation}
In the limit $k \to 0$, the only surviving term exactly matches the 2d IFT~\eqref{eq:2dift} we derived from 4d hBF theory.

\subsection{Reductions to 3d BF}

In the context of this example setup, we will now explicitly compute the reduction of 4d hBF to 3d BF, as well as the associated reduction of the 2d holomorphic IFT to a 1d IM. For comparison, we will also present the reduction of 4d CS to 3d BF, together with the reduction of its associated 2d IFT to a 1d IM. In particular, we will show that the 1d IM that comes from reducing the 2d holomorphic IFT also arises as a limit of the 1d IM that comes from reducing the 2d IFT constructed from 4d CS.

Let us perform a translational reduction along spacetime generated by the vector field $V = V^1 \pd_1 + V^2 \pd_2$ whose coefficients $V^1$ and $V^2$ are constant. We start by imposing the constraints $\LieD_V A = 0$ and $\LieD_V b = 0$ on the fundamental fields of 4d hBF theory, and then we contract $V$ into the $4$-form Lagrangian to find the $3$-form Lagrangian of 3d BF theory. The quotient space can be parametrised by the invariant coordinate $t = V^2 \, x^1 - V^1 \, x^2$, which satisfies $V(t) = 0$.

To simplify the contraction of $V$ into the Lagrangian, it is helpful to use the $(1,0)$-form shift symmetry to impose the condition $V \contract A = 0$. Solving this condition for the components of $\sigma$ in the transformation $A \mapsto A + \sigma_z \dr z + \sigma_w \dr w$, we find the new field configuration
\begin{equation}
A + \sigma_z \dr z + \sigma_w \dr w = -\frac{A_{\hat{w}}}{V^1 + z \, V^2} \, \dr t ~.
\end{equation}
In particular, we highlight that the reduction has introduced a singularity into the field configuration whose location depends on the choice of reduction vector $V$. While this might initially appear problematic, the meromorphic $(1,0)$-form of 3d BF theory will develop a complementary zero at this point ensuring that the whole setup is consistent. To see this explicitly, we recall the relationship $\tilde{\omega}_1 = V \contract \omega_2$ from which we compute
\begin{equation}
\tilde{\omega}_1 = - (V^1 + z \, V^2) \, \varphi(z) \, \dr z ~.
\end{equation}
As stated, the meromorphic $(1,0)$-form of 3d BF theory has an extra zero relative to the $(2,0)$-form of 4d hBF theory. This is analogous to the derivation of 4d CS as a reduction of 6d holomorphic CS theory \cite{Bittleston:2020hfv}.

To be precise, taking the definition $\tilde{\omega}_1 = V \contract \omega_2$ at face value would imply that $\tilde{\omega}_1$ is valued in $\cO(1)$ since we chose $\omega_2$ to be valued in $\cO(1)$ in our 4d hBF setup. Given that $b$ is a section of $\cO(-1)$ this is a perfectly valid choice.
However, we can remove the weight from $\tilde{\omega}_1$ in order to compare with the literature. To achieve this, we instead take the expression $\tilde{\omega}_1 = - (V^1 + z \, V^2) \, \varphi(z) \, \dr z$ at face value, effectively introducing an additional pole in $\tilde{\omega}_1$ at $z = \infty$. We compensate for this additional pole by introducing a complementary zero in $b$, or equivalently imposing the boundary condition $b \vert_\infty = 0$. In fact, we could have introduced this extra pole and zero at any point in $\CP^1$ without changing the theory, as we will discuss further in \secref{sec:minimal_setup}.

This 3d BF setup also inherits boundary conditions and residue conditions from the 4d hBF setup. The explicit expression for the 3d BF gauge field in terms of the 4d hBF gauge field shows that the 3d gauge field will satisfy the boundary conditions $A_t \vert_{z_p} = 0$ at each simple pole. In addition, the residue conditions $\res_\gamma (b) = q_\gamma \cdot \res_\gamma (A_{\hat{w}})$ in the 4d theory translate to
\begin{equation}
\res_\gamma (b) = -(V^1 + \gamma \, V^2) \, q_\gamma \cdot \res_\gamma (A_t) ~, \qquad
\res_{-V^1 / V^2} (b) = 0 ~.
\end{equation}
The second of these conditions comes from the observation that only $A_t$ picks up a pole in the reduction procedure. This completes the definition of the 3d BF setup.

Applying this reduction directly to the 2d theory~\eqref{eq:2dift} follows a similar logic; we impose that the fields are invariant under $V$ and then contract the vector field into the $2$-form Lagrangian. Under a generic reduction, we find an action of the form
\begin{equation}\label{eq:1diftsimple}
\Act_{\text{IM}_1}[g,\tilde g] = -\frac{q_\gamma (\alpha -\gamma)(\tilde\alpha-\gamma)}{(\alpha-\tilde\alpha)^2} \int_{\Real} \dr t \, \bl{c_\alpha \, \pd_t g g^{-1} - c_{\tilde \alpha} \, \pd_t \tilde{g} \tilde{g}^{-1}}{c_\alpha \, \pd_t g g^{-1} - c_{\tilde \alpha} \, \pd_t \tilde{g} \tilde{g}^{-1}} ~.
\end{equation}
The coefficients $c_\alpha = V^1 + \alpha \, V^2$ and $c_{\tilde \alpha} = V^1 + \tilde\alpha \, V^2$ depend on both the locations of the poles and on the choice of reduction vector. Let us also denote $c_\beta = V^2$. There are three special points in parameter space where either $c_\alpha = 0$ or $c_{\tilde \alpha} = 0$ or $c_\beta = 0$ (which implies $c_\alpha = c_{\tilde \alpha}$). Each of these cases corresponds to a reduction in the numbers of degrees of freedom as a consequence of the new zero colliding with one of the poles of $\varphi(z)$. In the three cases, the zero collides with $z = \alpha$ or $z = \tilde\alpha$ or $z = \infty$ respectively.

The 1d action~\eqref{eq:1diftsimple} has a gauge symmetry and a non-invertible kinetic term in the naive perturbative expansion, just as its 2d counterpart~\eqref{eq:2dift} did.
We can also arrive at~\eqref{eq:1diftsimple} by first reducing the 2d IFT found from 4d CS~\eqref{eq:4dcssimple} to give
\begin{equation}\begin{aligned}\label{eq:1diftinv}
\Act_{\text{IM}_1}[g,\tilde g] = & - k r_\alpha c_\alpha c_\beta \int_{\Real} \dr t \, \bl{g^{-1}\pd_t g}{g^{-1} \pd_t g}
- k r_{\tilde{\alpha}} c_{\tilde{\alpha}} c_\beta \int_{\Real} \dr t \, \bl{\tilde g^{-1}\pd_t \tilde g}{\tilde g^{-1} \pd_t\tilde g}
\\
& \hspace{-2em} - \frac{(\alpha - \gamma) (\tilde{\alpha} - \gamma)}{(\alpha - \tilde{\alpha})^2} \bigg( q_\gamma - \frac{k\bar{\gamma}}{1 + \gamma \bar{\gamma}} \bigg) \int_{\Real} \dr t \, \bl{c_\alpha \, \pd_t g g^{-1} - c_{\tilde \alpha} \, \pd_t \tilde{g} \tilde{g}^{-1}}{c_\alpha \, \pd_t g g^{-1} - c_{\tilde \alpha} \, \pd_t \tilde{g} \tilde{g}^{-1}} ~.
\end{aligned}
\end{equation}
and then taking $k\to0$.
This intermediate theory has an invertible kinetic term.
One approach to quantizing the models with a non-invertible kinetic term~\eqref{eq:1diftsimple} and~\eqref{eq:2dift} may be to start by quantizing their invertible counterparts~\eqref{eq:1diftinv} and~\eqref{eq:4dcssimple} and then studying the limit in the quantum theory.

The 1d action~\eqref{eq:1diftinv} can also be derived directly from 3d BF, and the appropriate setup can be determined by applying the reduction to 4d CS. To ease the comparison of this setup with the 3d BF setup we derived from 4d hBF, we will apply a few modifications that do not change the underlying 1d model (these freedoms are discussed in \secref{sec:minimal_setup}). The setup we recover as a reduction from 4d CS has one simple zero and three simple poles, and all of the boundary and residue conditions impose mixed constraints on $A_t$ and $b$. We will turn the simple pole at $z = \infty$ into a double pole and introduce an additional zero at $z = \tilde\gamma$ (to match with the parameters above we take $V^2 = 1$ and $V^1 = -\tilde\gamma$). Complementary to this new pole and zero, we will introduce a zero and pole in $b$ at the same points, i.e.~$b \vert_\infty = 0$ and $\res_{\tilde\gamma} (b) \neq 0$. Then, we make use of a shift symmetry (discussed in \secref{sec:minimal_setup}) to simplify the boundary conditions in exchange for mixed residue conditions.

In summary, the setup we land on has two simple poles at $z = \alpha$ and $z = \tilde\alpha$, a double pole at $z = \infty$, and two simple zeroes at $z = \gamma$ and $z = \tilde\gamma$ in the $(1,0)$-form $\tilde{\omega}_1 = \frac{(z - \gamma)(\tilde\gamma - z)}{(z - \alpha)(z - \tilde\alpha)} \dr z$. The boundary conditions are $A \vert_\infty = 0$ and $b\vert_\infty = 0$ at the double pole, and $A\vert_\alpha = 0$ and $A\vert_{\tilde\alpha} = 0$ at the simple poles. Both $A$ and $b$ are allowed to have simple poles at $z = \gamma$ and $z = \tilde\gamma$, and their residues must satisfy $\res_\gamma (b) = \mathfrak{q}_\gamma \res_\gamma (A_t)$ and $\res_{\tilde\gamma} (b) = \mathfrak{q}_{\tilde\gamma} \res_{\tilde\gamma} (A_t)$. The 3d BF theory action is given by
\begin{equation}
\Act_{\text{BF}_3} = \frac{1}{2 \pi \rmi} \int_{\CP^1 \times \Real} \tilde{\omega}_1 \wedge \bl{b}{F}~.
\end{equation}
The on-shell configurations are
\begin{equation}\begin{aligned}
A_t & = -\frac{(z - \tilde\alpha)(\alpha - \gamma)(\alpha - \tilde\gamma) \pd_t g g^{-1} - (z - \alpha)(\tilde\alpha - \gamma)(\tilde\alpha - \tilde\gamma) \pd_t \tilde{g} \tilde{g}^{-1}}{(\alpha - \tilde\alpha)(z - \gamma)(z - \tilde\gamma)} ~, \\
b & = \mathfrak{q}_\gamma \frac{\res_\gamma (A_t)}{z - \gamma} + \mathfrak{q}_{\tilde\gamma} \frac{\res_{\tilde\gamma} (A_t)}{z - \tilde\gamma} ~.
\end{aligned}\end{equation}
Substituting these into the action gives the 1d theory
\begin{equation}\begin{aligned}
\Act_{\text{IM}_1} = & \frac{\mathfrak{q}_\gamma (\alpha-\gamma)(\tilde\alpha-\gamma)}{(\gamma - \tilde\gamma)(\alpha-\tilde\alpha)^2} \int_{\Real} \dr t \, \bl{c_\alpha \pd_t g g^{-1} - c_{\tilde\alpha} \pd_t \tilde{g} \tilde{g}^{-1}}{c_\alpha \pd_t g g^{-1} - c_{\tilde\alpha} \pd_t \tilde{g} \tilde{g}^{-1}} \\
& - \frac{\mathfrak{q}_{\tilde\gamma} (\alpha-\tilde\gamma)(\tilde\alpha-\tilde\gamma)}{(\gamma - \tilde\gamma)(\alpha-\tilde\alpha)^2} \int_{\Real} \dr t \, \bl{\tilde c_\alpha \pd_t g g^{-1} + \tilde c_{\tilde\alpha} \pd_t \tilde{g} \tilde{g}^{-1}}{\tilde c_\alpha \pd_t g g^{-1} + \tilde c_{\tilde\alpha}\pd_t \tilde{g} \tilde{g}^{-1}} ~, \\
\end{aligned}\end{equation}
where we have defined $\tilde c_\alpha = \alpha - \gamma$ and $\tilde c_{\tilde\alpha} = \tilde\alpha-\gamma$.
This matches eq.~\eqref{eq:1diftsimple} in the limit $\mathfrak{q}_\gamma \to -(\gamma - \tilde\gamma) q_\gamma$ and $\mathfrak{q}_{\tilde\gamma} \to 0$, which also recovers the residue conditions derived from 4d hBF.

\subsection{On the minimal setup}\label{sec:minimal_setup}

Having worked through a concrete example, let us return to more general features of 4d hBF theory and discuss some finer details that we have thus far neglected. In particular, we would like to address the question of whether or not this setup is the minimal or simplest setup accessible and justify some of the choices we have made along the way.

When defining 4d hBF theory, we decided to give $\omega_2$ and $b$ non-vanishing weights, such that they are valued in non-trivial line bundles over $\CP^1$. Since these objects only appear in the action through their product $B = \omega_2 \otimes b \in \Omega^{2,0}(M, \alg{g})$, there is some freedom in how we choose to distribute weights between them. Nonetheless, it is worth highlighting that working exclusively in terms of $B$ appears\footnote{It may be possible to encode this data into $B$ by allowing it to have poles at fixed points and imposing boundary conditions on $A$ at these points. We will not pursue this possibility here.} to lead to a different theory, since it forgets about the non-dynamical data $\varphi(z)$, which plays an important role in our constructions.

Let us consider a general setup, along the lines discussed above, in which $\omega_2$ is valued in $\cO(1)$ and $b$ is valued in $\cO(-1)$. We can trade this setup for an equivalent setup in which both $\omega_2$ and $b$ are valued in trivial line bundles by making use of the following symmetry. If we pick an arbitrary point $\rho \in \CP^1$, we can trade\footnote{This is most clear in homogeneous coordinates where we are proposing the transformation $\omega_2 \mapsto \omega_2 / \langle \lambda \rho \rangle$ and $b \mapsto \langle \lambda \rho \rangle b$.} any section of $\cO(1)$ for a section of the trivial bundle with a singularity at $z = \rho$, and any section of $\cO(-1)$ for a section of the trivial bundle that vanishes at $z = \rho$. This amounts to creating a new pole in $\omega_2$ and imposing the boundary condition that $b$ must vanish at this pole.

At first sight, it is not clear that this alternative setup should be equivalent since there is a new pole that might source new dynamical field content in the 2d theory. To resolve this, we recall that the dynamical field content of the 2d theory parametrises the gauge symmetry that is broken by the boundary conditions. In this example, the boundary condition $b \vert_{z = \rho} = 0$ does not break any of the gauge symmetries of the theory, hence does not give rise to any additional degrees of freedom. Explicit calculation verifies that the 2d theory is unchanged by this exchange of weights between $\omega_2$ and $b$.

\medskip

Another feature worth discussing is the unusual residue conditions $\res_\gamma (b) = q_\gamma \cdot \res_\gamma (A_{\hat{w}})$ that we imposed on the fields at the zeroes of $\omega_2$. This is in contrast to the rather simple boundary conditions $A_{\hat{w}} \vert_{z = z_p} = 0$ we imposed, and arguably in contrast to the more typical residue conditions found in the 4d CS literature, along the lines of $\res_\gamma (A_+) = 0$ and $\res_{\tilde \gamma}(A_-) = 0$ in lightcone coordinates. We will argue that 4d hBF theory has a shift symmetry that allows us to trade the unusual residue conditions for more familiar ones, at the expense of complicating the boundary condition. It is then a matter of preference in which framework we choose to work.

Let us recall the action of 4d hBF theory,
\begin{equation}
\Act_{\text{hBF}_4} = \frac{1}{2 \pi \rmi} \int \omega_2 \wedge \bl{b}{F(A)} + \frac{1}{4 \pi \rmi} \int \dr \omega_2 \wedge \bl{b}{A} ~.
\end{equation}
In this expression, we have included the boundary term that we derived from the limit of 4d CS theory. It turns out that this boundary term, with this particular coefficient, leads to the action being invariant under the shift symmetry
\begin{equation}
A_{\hat{w}} \mapsto A_{\hat{w}} - \xi \, b ~, \qquad
\pd_{\bar{z}} \xi = 0 ~, \qquad
\xi \in C^\infty(M) ~.
\end{equation}
This transformation is always a symmetry at the level of the equations of motion, since it preserves both $\pd_{\bar{z}} A_{\hat{w}} = 0$ and the integrability condition $\pd_{\hat{w}} b + [A_{\hat{w}} , b] = 0$, and this specific boundary term ensures that it also leaves the action invariant.

For the moment, the relevance of this shift symmetry is that we can use it to remove any poles in $A_{\hat{w}}$ that satisfy the residue condition $\res_\gamma (b) = q_\gamma \cdot \res_\gamma (A_{\hat{w}})$. This implies that there should be an equivalent setup in which only $b$ is allowed poles at the zeroes of $\omega_2$ and we have $\res_\gamma (A_{\hat{w}}) = 0$. Applying the transformation above and solving for $\xi$ to remove the poles from $A_{\hat{w}}$, we will find a new field configuration that solves the equations of motion. On the other hand, this transformation does not preserve the boundary conditions.
Instead, the new setup will come with boundary conditions of the form $A_{\hat{w}} \vert_{z = z_p} = \xi \, b \vert_{z = z_p}$. In this sense, we are able to trade the simple boundary conditions for simple residue conditions, and vice versa, but in either case one of these constraints must relate $A_{\hat{w}}$ and $b$.

It is important to highlight that our previous analysis is not modified by the inclusion of the boundary term in the 4d hBF theory action. Since we have worked with the boundary conditions $A_{\hat{w}} \vert_{z = z_p} = 0$, this boundary term vanishes on-shell and does not contribute to the action of the 2d theory. By comparison, had we worked with the equivalent setup in which $\res_\gamma (A_{\hat{w}}) = 0$ and $A_{\hat{w}} \vert_{z = z_p} = \xi \, b \vert_{z = z_p}$, it would be the bulk term in the action that did not contribute and the boundary term would have sourced the entire action of the 2d theory. Since both of these terms may be relevant in a generic setup, it is natural to include the boundary term in the definition of 4d hBF theory.

\medskip

Having addressed these two finer details of 4d hBF theory, let us consider the minimal or simplest setup. We will argue that the example in the previous section is one of the simplest setups available. Whatever the 2d theory that emerges from a given 4d hBF theory setup, we expect its equations of motion to be equivalent to the integrability condition $\pd_{\hat{w}} b + [A_{\hat{w}} , b] = 0$. While this equation may be non-trivial if $A_{\hat{w}}$ vanishes on-shell, if $b$ is identically zero (once we have solved $\pd_{\bar{z}} b = 0$ and imposed the boundary conditions) then the 2d theory will have a trivial equation of motion. In order to avoid this circumstance, we must ensure that the on-shell field configuration for $b$ is non-vanishing.

Let us consider the case that $b$ and $\omega_2$ have no homogeneous weights, so that they live in the trivial bundle over $\CP^1$ rather than $\cO(-1)$ and $\cO(1)$ respectively. The total number of poles of $\omega_2$ minus the total number of zeroes (counted with multiplicity) will always be three, and we will say that $\omega_2$ has no zeroes in a minimal setup. Since $\omega_2$ has no zeroes, we do not allow our field configurations to have singularities, and the solution to $\pd_{\bar{z}} b^\prime = 0$ is that $b^\prime$ must be constant in the $z$-plane. This rules out the boundary condition $b \vert_{z = z_p} = 0$ as it would imply that $b$ vanishes everywhere in the $z$-plane, leading to a trivial theory.

At a simple pole of $\omega_2$, the two simplest boundary conditions\footnote{The shift symmetry argument shows that any boundary condition of the form $A_{\hat{w}} \vert_{z = z_p} = \xi \, b \vert_{z = z_p}$ can be brought to the boundary condition $A_{\hat{w}} \vert_{z = z_p} = 0$, at the expense of modifying the residue conditions on the fields.} we could consider are $b \vert_{z = z_p} = 0$ or $A_{\hat{w}} \vert_{z = z_p} = 0$. There is a key difference between these two constraints, namely the condition $A_{\hat{w}} \vert_{z = z_p} = 0$ breaks the gauge symmetry at this pole, while the condition $b \vert_{z = z_p} = 0$ does not. Since the degrees of freedom of the 2d IFT arise from the edge mode field $\hat{g}$ and parametrise the broken gauge symmetry at the poles, this changes the number of fields in the 2d theory. If we impose the condition $A_{\hat{w}} \vert_{z = z_p} = 0$, then the pole will source a degree of freedom, while this degree of freedom will be gauge-trivial if we impose $b \vert_{z = z_p} = 0$.

We have excluded the boundary condition $b \vert_{z = z_p} = 0$ as it would lead to $b = 0$ everywhere, so we are led to impose $A_{\hat{w}} \vert_{z = z_p} = 0$ at each simple pole. A priori, this leads to three group-valued degrees of freedom in the minimal setup, one for each simple pole of $\omega_2$. However, the gauge symmetries parametrised by $\check{h}$ constant in the $z$-plane are sufficient to gauge fix one group-valued degree of freedom. We therefore expect the minimal setup to be a 2d IFT with two group-valued degrees of freedom, and the previous section presents an example of one such theory. This analysis is not fundamentally changed by considering higher-order poles in $\omega_2$, though group-valued degrees of freedom will be traded for algebra-valued degrees of freedom, as we present in~\secref{sec:doublepole}.

\subsection{Double pole setup}\label{sec:doublepole}

We can also consider 4d hBF theory with alternative defect setups. In the context of disorder defects, this corresponds to different choices of $\omega_2$, the meromorphic $(2,0)$-form. In this section, we will explore another simple setup, where $\omega_2$ has a finite double pole at $z = \alpha$ and a simple pole at $z = \infty$. We take $b$ to be valued in $\cO(-1)$ and $\omega_2$ to be valued in $\cO(1)$.

Explicitly, the meromorphic $(2,0)$-form\footnote{In homogeneous coordinates, this $(2,0)$-form is written as $\omega_2 = \frac{\langle \lambda \gamma \rangle}{\langle \lambda \alpha \rangle^2 \langle \lambda \beta \rangle} \langle \lambda \dr \lambda \rangle \wedge \lambda_a \dr x^a$.} is given by
\begin{equation}
\omega_2 = \frac{z - \gamma}{(z - \alpha)^2} \, \dr z \wedge \dr w ~.
\end{equation}
Since this example is not the main focus of this paper we will highlight the key results rather than providing a detailed derivation. At the poles of $\omega_2$, we impose the boundary conditions
\begin{equation}
A_{\hat{w}} e^{\hat{w}} \vert_{z = \infty} = 0 ~, \qquad
A_{\hat{w}} \vert_{z = \alpha} = 0 ~, \qquad
\pd_z A_{\hat{w}} \vert_{z = \alpha} = 0 ~.
\end{equation}
In addition, we will allow both $A_{\hat{w}}$ and $b$ to have a simple pole at $z = \gamma$, subject to the residue condition $\res_\gamma (b) = q_\gamma \cdot \res_\gamma (A_{\hat{w}})$.

We introduce the edge mode field $\hat{g}$ via the field redefinitions $A = \hat{g}^{-1} A^\prime \hat{g} + \hat{g}^{-1} \dr \hat{g}$ and $b = \hat{g}^{-1} b^\prime \hat{g}$, and there is sufficient gauge freedom to bring the relevant degrees of freedom to the form
\begin{equation}
\hat{g} \vert_\infty = \id ~, \qquad
\hat{g} \vert_\alpha = g ~, \qquad
\hat{g}^{-1} \pd_z \hat{g} \vert_\alpha = u ~.
\end{equation}
Solving the equations of motion $\varphi \cdot \pd_{\bar{z}} A_{\hat{w}}^\prime = 0$ and $\varphi \cdot \pd_{\bar{z}} b^\prime = 0$ in the gauge $A_{\bar{z}}^\prime = 0$, and then imposing the boundary conditions, we find the on-shell field configurations
\begin{equation}\begin{aligned}
A_{\hat{w}}^\prime & = -\pd_\alpha g g^{-1} - \frac{(z - \alpha)(\alpha - \gamma)}{z - \gamma} \big( \pd_\beta g g^{-1} + g \cdot \pd_\alpha u \cdot g^{-1} \big) ~, \\
b^\prime & = q_\gamma \frac{(\alpha - \gamma)^2}{z - \gamma} \big( \pd_\beta g g^{-1} + g \cdot \pd_\alpha u \cdot g^{-1} \big) ~.
\end{aligned}\end{equation}
Substituting these on-shell field configurations into the action gives the 2d theory
\begin{equation}\label{eq:2diftdouble}
\Act_{\text{IFT}_2}[g,u] = q_\gamma \, (\alpha - \gamma)^2 \int_{\Real^2} \dr^2 x \, \bl{g^{-1} \pd_\beta g + \pd_\alpha u}{g^{-1} \pd_\beta g + \pd_\alpha u} ~,
\end{equation}
whose equations of motion are equivalent to the integrability condition $\pd_{\hat{w}} b^\prime + [A^\prime_{\hat{w}}, b^\prime] = 0$.

To recover this 2d action from 4d CS theory, we start with the meromorphic $(1,0)$-form
\begin{equation}
\omega_1 = \frac{z - \gamma}{(z - \alpha)^2} \dr z ~.
\end{equation}
Working in the basis of 1-forms $\{ e^w, e^{\hat{w}} \}$, we impose the boundary conditions
\begin{equation}
A_{\hat{w}} e^{\hat{w}} \vert_{z = \infty} = 0 ~, \qquad
A_{\hat{w}} \vert_{z = \alpha} = 0 ~, \qquad
\pd_z A_{\hat{w}} \vert_{z = \alpha} = 0 ~.
\end{equation}
In addition, we allow both $A_w$ and $A_{\hat{w}}$ to have a simple pole at $z = \gamma$, subject to the residue condition $\res_\gamma (k \, A_{w}) = q_\gamma \cdot \res_\gamma (A_{\hat{w}})$. To compute the 2d action, we introduce the edge mode field via the field redefinition $A = \hat{g}^{-1} A^\prime \hat{g} + \hat{g}^{-1} \dr \hat{g}$ and impose the constraints $A^\prime_{\bar{z}} = 0$ and $\hat{g} \vert_\infty = \id$. Solving the equations of motion that determine the $z$-dependence, and then solving the boundary conditions for $A^\prime$ in terms of $g = \hat{g} \vert_\alpha$ and $u = \hat{g}^{-1} \pd_z \hat{g} \vert_\alpha$ gives the on-shell field configurations
\begin{equation}\begin{aligned}
A^\prime_{\hat{w}} & = -\pd_\alpha g g^{-1} - \frac{(z - \alpha)(\alpha - \gamma)}{z - \gamma} \big( \pd_\beta g g^{-1} + g \cdot \pd_\alpha u \cdot g^{-1} \big) ~, \\
A^\prime_w & = \frac{\bar{z}}{1 + z \bar{z}} A^\prime_{\hat{w}} + \bigg( \frac{q_\gamma}{k} - \frac{\bar{\gamma}}{1 + \gamma \bar{\gamma}} \bigg) \frac{\res_\gamma (A^\prime_{\hat{w}})}{z - \gamma} ~.
\end{aligned}\end{equation}
These match the on-shell field configurations of the 4d hBF setup in the limit $k \to 0$. Substituting these expressions into the 4d CS action, we can compute the 2d action
\begin{equation}\begin{aligned}\label{eq:4dcsdouble}
\Act_{\text{IFT}_2} [g, u] & = k \, \Act_{\text{WZW}_{\alpha \beta}} [g]
-2 k \, (\alpha - \gamma)\int_{\Real^2} \dr^2 x \,\bl{g^{-1}\pd_\beta g}{\pd_\alpha u}
- k \, (\alpha - \gamma) \int_{\Real^2} \dr^2 x \, \bl{g^{-1} \pd_\beta g }{g^{-1} \pd_\beta g } \\
& \qquad + (\alpha - \gamma)^2 \bigg( q_\gamma - \frac{k \, \bar{\gamma}}{1 + \gamma \bar{\gamma}} \bigg) \int_{\Real^2} \dr^2 x \, \bl{g^{-1} \pd_\beta g + \pd_\alpha u}{g^{-1} \pd_\beta g + \pd_\alpha u} ~.
\end{aligned}\end{equation}
The first two terms can be understood as a WZW model on $\mathsf{T}^*\grp{G}$.
The 2d IFT~\eqref{eq:4dcsdouble} is invariant under the symmetries
\begin{equation}\begin{gathered}
g \mapsto h_\beta^{-1} \cdot g \cdot h_\alpha ~, \qquad
u \mapsto h_\alpha^{-1} \cdot u \cdot h_\alpha + v_\alpha ~, \\
\dr h_\alpha = 0 ~, \qquad
\pd_\beta h_\beta = 0 ~, \qquad
\pd_\alpha v_\alpha = 0 ~.
\end{gathered}\end{equation}
In the limit $k \to 0$, the only surviving term matches the 2d action derived from 4d hBF theory.

The 2d theory~\eqref{eq:2diftdouble} can also be found as the $\tilde\alpha \to \alpha$ limit of the 2d theory~\eqref{eq:2dift}.
Explicitly, we set
\begin{equation}
\tilde g = g e^{-(\alpha-\tilde \alpha) u} ~.
\end{equation}
Recalling that $\pd_{\tilde\alpha} = \pd_\alpha - (\alpha-\tilde\alpha)\pd_\beta$ we find
\begin{equation}
(\pd_\alpha g g^{-1} - \pd_{\tilde\alpha} \tilde g \tilde g^{-1}) = (\alpha-\tilde\alpha) (\pd_\beta g g^{-1} + g
\cdot \pd_\alpha u \cdot g^{-1}) + \mathcal{O} \big( (\alpha-\tilde\alpha)^2 \big) ~,
\end{equation}
and immediately see that that the action~\eqref{eq:2diftdouble} is the $\tilde\alpha \to \alpha$ limit of the action~\eqref{eq:2dift}.
We can also take the same limit in the action~\eqref{eq:4dcssimple} yielding~\eqref{eq:4dcsdouble}.

\subsection{Order defect setup}\label{sec:order}

As our final example, let us consider a setup in 4d hBF theory with order defects. As before, we will take the field $b$ to be valued in $\cO(-1)$ and the meromorphic $(2,0)$-form to be valued in $\cO(1)$. This time, we consider\footnote{In homogeneous coordinates, this $(2,0)$-form is $\omega_2 = \frac{1}{\langle \lambda \alpha \rangle \langle \lambda \beta \rangle} \langle \lambda \dr \lambda \rangle \wedge \lambda_a \dr x^a$.} the $(2,0)$-form
\begin{equation}
\omega_2 = \frac{1}{z - \alpha} \, \dr z \wedge \dr w ~.
\end{equation}
This has a simple pole at $z = \alpha$, a simple pole at $z = \infty$, and it is nowhere vanishing. At each of these simple poles $z = z_p$, we will impose the boundary condition $A_{\hat{w}} \vert_{z = z_p} = 0$.

As it is, this setup would lead to a trivial theory in two-dimensions. The reason for this is that we have not allowed $b$ to have any singularities in the $z$-plane, so the only solution to $\pd_{\bar{z}} b = 0$ is for $b$ itself to vanish, $b = 0$. Even though the field configuration for $A$ would be non-vanishing, the action is linear in $b$ so it would also vanish once we substitute in $b = 0$. We will rectify this situation by introducing order defects into our setup.

Rather than simply considering the action $\int \omega_2 \wedge \bl{b}{F}$, we will add two order defects to the action such that it takes the form
\begin{equation}
\Act_{\text{hBF}_4} = \frac{1}{2 \pi \rmi} \int \omega_2 \wedge \bl{b}{F(A)}
+ \int_{\{ z = \eta \}} \dr w \wedge \bl{h^{-1} \phi h}{h^{-1} \dr h + h^{-1} A h}
+ \int_{\{ z = \kappa \}} \dr^2 x \, \mathsf{P}(b) ~.
\end{equation}
The two additional terms we have added live at fixed points $z = \eta$ and $z = \kappa$ in the $z$-plane. The first new term hosts additional degrees of freedom $\phi \in C^\infty (\Real^2, \alg{g})$ and $h \in C^\infty(\Real^2, \grp{G})$, which live on the defect, while the second term is built from an ad-invariant polynomial on the algebra $\mathsf{P} : \alg{g} \to \Complex$. Each of these terms will introduce new contributions to the bulk equations of motion, which will in turn source singularities in the on-shell field configurations.

We can see this explicitly by introducing the field redefinitions $A \mapsto \hat{g}^{-1} A^\prime \hat{g} + \hat{g}^{-1} \dr \hat{g}$, $b \mapsto \hat{g}^{-1} b^\prime \hat{g}$, $h \mapsto \hat{g}^{-1} h$ and $\phi \mapsto \hat{g}^{-1} \phi \hat{g}$. The action is invariant under gauge transformations so it does not change, but the boundary conditions change when written in terms of the new field content. Most importantly, this allows us to gauge fix $A^\prime_{\bar{z}} = 0$. After imposing this constraint, there is still sufficient residual gauge symmetry to impose $\hat{g} \vert_\infty = \id$, and we will denote the evaluation of the edge mode field at the other boundary by $\hat{g} \vert_\alpha = g$. The equations of motion for $A^\prime$ and $b^\prime$ are given by
\begin{equation}\begin{aligned}
\varphi \, \pd_{\bar{z}} A^\prime_{\hat{w}} & = - 2 \pi \rmi \, \delta^{(2)}(z - \kappa) \, \mathsf{P}^\prime (b^\prime) ~, \\
\varphi \, \pd_{\bar{z}} b^\prime & = - 2 \pi \rmi \, \delta^{(2)}(z - \eta) \, \phi ~.
\end{aligned}\end{equation}
In the first equation, $\mathsf{P}^\prime (b^\prime)$ is defined by the relation $\delta \mathsf{P} (b^\prime) = \bl{\delta b^\prime}{\mathsf{P}^\prime (b^\prime)}$.

The localised terms that come from the variations of the order defects have the effect of sourcing singularities in the on-shell field configurations with fixed residues. Taking into account the global behaviour of these fields, the most general solutions\footnote{In homogeneous coordinates, these solutions are $A^\prime_{\hat{w}} = \langle \alpha \lambda \rangle \, a_\infty + \langle \lambda \beta \rangle \, a_\alpha + \frac{\langle \lambda \beta \rangle \langle \lambda \alpha \rangle}{\langle \lambda \kappa \rangle} \mathsf{P}^\prime (b \vert_\kappa)$ and $b^\prime = \frac{\langle \alpha \eta \rangle}{\langle \lambda \eta \rangle} \phi$.} are given by
\begin{equation}
A^\prime_{\hat{w}} = z \, a_\infty + a_\alpha + \frac{z - \alpha}{z - \kappa} \, \mathsf{P}^\prime (b^\prime \vert_\kappa) ~, \qquad
b^\prime = \frac{\alpha - \eta}{z - \eta} \phi ~.
\end{equation}
We can now substitute these solutions into the boundary conditions to solve for the remaining unknowns. This gives the field configurations
\begin{equation}
A^\prime_{\hat{w}} = -\pd_\alpha g g^{-1} + \frac{z - \alpha}{z - \kappa} \, \mathsf{P}^\prime (b^\prime \vert_\kappa) ~, \qquad
b^\prime = \frac{\alpha - \eta}{z - \eta} \phi ~.
\end{equation}
The final step is to substitute these configurations back into the action to find the 2d integrable theory. Including both the contributions from the bulk action and from the order defects, we are left with the 2d action
\begin{equation}
\Act_{\text{IFT}_2} = \int \dr^2 x \, \bl{\phi}{\pd_\eta h h^{-1} - \pd_\alpha g g^{-1}} + \int \dr^2 x \, \mathsf{P} \big( \tfrac{\alpha - \eta}{\kappa - \eta} \phi \big) ~.
\end{equation}
If we make the simple choice $\mathsf{P}(X) = \frac{1}{2} \bl{X}{X}$ then we can integrate out $\phi$ to recover a 2d IFT with two group-valued fields. In fact, we find the same 2d IFT as the one we recovered from disorder defects~\eqref{eq:2dift} if we make the identification $\pd_\eta h h^{-1} = \pd_{\tilde\alpha} \tilde{g} \tilde{g}^{-1}$ and suitably relate the parameters.

Now let us consider the reduction to 3d BF.
This setup has a simple pole at $z = \alpha$ and a simple zero at $z = \gamma$ in the $(1,0)$-form $\omega = \frac{z - \gamma}{z - \alpha} \dr z$.
Employing the terminology of~\cite{Vicedo:2022mrm}, it also has one `type-A' order defect at $z = \eta$ and one `type-B' order defect at $z = \kappa$. We impose the boundary conditions $A\vert_\infty = 0$ and $b\vert_\infty = 0$ and gauge fix the edge modes at this pole. We also impose the boundary condition $A\vert_\alpha = 0$ and denote the edge mode at this pole by $g$. We allow both $A$ and $b$ to have a simple pole at $z = \gamma$ and demand that the residues satisfy $\res_\gamma (b) = 0$. The 3d BF theory action is given by
\begin{equation}
\Act_{\text{BF}_3} = \frac{1}{2 \pi \rmi} \int_{\CP^1 \times \Real} \omega \wedge \bl{b}{F} - \int_{\{\eta\} \times \Real} \bl{h^{-1} \phi h}{h^{-1} \dr h + h^{-1} A h} + \int_{\{\kappa\} \times \Real} \dr t \, \mathsf{P}(b) ~.
\end{equation}
The on-shell configurations are
\begin{equation}
A_t = \frac{(\gamma - \alpha) \pd_t g g^{-1}}{z - \gamma} + \frac{(z - \alpha) \mathsf{P}^\prime(b\vert_\kappa)}{(z - \gamma)(z - \kappa)} ~, \qquad
b = \frac{(\eta - \alpha) \, \phi}{(\eta - \gamma)(z - \eta)} ~.
\end{equation}
Substituting these into the action gives the 1d theory
\begin{equation}
\Act_{\text{IM}_1} = \int_{\Real} \dr t \, \bigg[ -\bl{\phi}{\pd_t h h^{-1}} + \frac{\gamma - \alpha}{\gamma - \eta} \bl{\phi}{\pd_t g g^{-1}} + \mathsf{P}(b\vert_\kappa) \bigg] ~.
\end{equation}
Taking either $\gamma \to \eta$ while rescaling $\phi$ or $\gamma \to \alpha$ we find a one field model that is equivalent to the one field model of~\cite{Vicedo:2022mrm}.
Moreover, the $\gamma \to \alpha$ limit explicitly gives the order defect setup considered there.
Alternatively, we can relax $\res_\gamma(b) = 0$ to get a two field model with an invertible kinetic term.
This model is equivalent to the two field model of \cite{Vicedo:2022mrm} also found solely from order defects.

\section{Concluding remarks}\label{sec:conc}

In this paper, we have investigated the network of theories
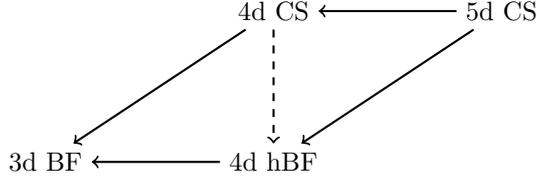
\begin{equation*}
\begin{tikzpicture}
\node (5mCS) at (9,1) {5d CS};
\node (4mCS) at (6,1) {4d CS};
\node (3mBF) at (3,-1) {3d BF};
\node (4hBF) at (6,-1) {4d hBF};
\draw[thick,->] (5mCS)--(4mCS);
\draw[thick,->] (5mCS)--(4hBF);
\draw[thick,->] (4mCS)--(3mBF);
\draw[thick,->] (4hBF)--(3mBF);
\draw[thick,dashed,->] (4mCS)--(4hBF);
\end{tikzpicture}
\end{equation*}
In \secref{sec:4dhbf}, we gave a precise meaning to each of the arrows in this diagram, thereby describing the relationships between these theories.
Having introduced 4d holomorphic BF, we explained how it appears as the $k\to0$ limit of 4d mixed Chern-Simons by taking one of the fields to scale with $1/k$.
We also presented the reductions from 4d hBF and 4d CS to 3d BF, and from 5d CS to 4d CS and 4d hBF completing the network.
In \secref{sec:defects}, we gave explicit examples of the relation between 4d hBF, 4d CS and 3d BF (together with their associated integrable systems) using some example defect setups.

An important step in this construction was to reframe the regularity conditions at disorder defects in terms of residue conditions on fields, enabling us to write the generalised residue conditions in eqs.~\eqref{eq:res1} and~\eqref{eq:res2}.
In particular, in contrast to taking the fields to be regular and specifying their pole structure, we allow the fields to have singularities at the disorder defects and specify their residues to ensure the finiteness of the action.
It may be possible to generalise further to incorporate free functions of the residues, potentially recovering the auxiliary field models of~\cite{Ferko:2024ali,Bielli:2024ach,Babaei-Aghbolagh:2025uoz,Babaei-Aghbolagh:2025hlm,Baglioni:2025tsc}.
\unskip\footnote{We thank C. Ferko, D. Bielli and N. Baglioni for discussions on this point.}
This method would be complementary to those proposed in \cite{Fukushima:2024nxm,Sakamoto:2025hwi,Fukushima:2025tlj} and it would be interesting to understand the relation between the different approaches.
Moreover, in the context of 4d hBF and 3d BF, it would also be interesting to see if the generalised residue conditions can be used to construct a disorder defect origin for the `type B' order defects discussed in \secref{sec:order}.

\medskip

One of the objectives of this paper was to compare 2d holomorphic integrability to the more standard notions of integrability in one and two dimensions.
Let us summarise the key results of this comparison.
Integrability in the 2d theories coming from 4d CS is described by a meromorphic Lax connection,
\begin{equation}
\pd_{\bar{z}} A_a = 0 ~, \qquad
\pd_a A_b - \pd_b A_a + [A_a, A_b] = 0 ~, \qquad
a \in \{1,2\} ~.
\end{equation}
These are the equations of motion of 4d CS~\cite{Costello:2019tri} in the gauge $A_{\bar{z}} = 0$.
Infinitely many conserved charges can be constructed by performing a Laurent expansion of the Wilson line along a spatial slice,
\begin{equation}
U = \Pexp \bigg( - \int \dr x \, A_x \bigg) ~, \qquad
U = \sum_{n=-\infty}^\infty z^n U_n ~.
\end{equation}
In addition to these non-local charges, there is a tower of local involutive higher spin charges~\cite{Lacroix:2017isl},
\begin{equation}
Q_n = \int \dr x \, \Tr \big( \res_{\gamma} (A_x)^n \big) ~,
\end{equation}
where $z = \gamma$ is a simple pole of $A_x$.
Similarly, the 3d BF equations of motion~\cite{Vicedo:2022mrm} in the gauge $A_{\bar z} = 0$ are
\begin{equation}
\pd_{\bar{z}} A_{t} = 0 ~, \qquad
\pd_{\bar{z}} b = 0 ~, \qquad
\pd_t b + [A_t, b] = 0 ~.
\end{equation}
The pair $(A_t, b)$ becomes the meromorphic Lax pair of the 1d IM, and the conserved charges are
\begin{equation}
\Tr \big( \res_{\gamma} (b)^n \big)
\end{equation}
From the 3d BF perspective, the Wilson line for $A$ is another gauge-invariant operator, but we are forced to integrate over the time direction so this does not produce a conserved charge in the 1d IM.

Taking the gauge $A_{\bar{z}} = 0$ in 4d hBF with $\mathcal{O}(1)$ as the underlying complex manifold, the analogous equations are
\begin{equation}\label{eq:4dholbfeom}
\pd_{\bar{z}} A_a = 0 ~, \qquad
\pd_{\bar{z}} b = 0 ~, \qquad
\pd_2 b - z \, \pd_1 b + [A_2, b] - z \, [A_1, b] = 0 ~.
\end{equation}
Only the combination $A_{\hat{w}} = A_2 - z A_1$ appears in the analogue of the Lax equation, so the pair $(A_{\hat{w}}, b)$ is sufficient to capture the integrability of the associated 2d theory.
The explicit dependence on $z$ in these equations is what results in the theory having non-trivial dynamics in 2d rather than 1d.
If we take the underlying complex manifold to be the trivial bundle $\Complex \times \CP^1$ (as considered in~\cite{Winstone:2023fpe}), we would find the same equations with $z$ replaced by $\rmi$, which is simply a complexification of 3d BF.

The natural operators in 4d hBF are
\begin{equation}
\Pexp \bigg( - \int_{\Complex} \dr^2 w \, A_{\hat{w}} \bigg) ~, \qquad
\int \dr x \, \Tr \big( \res_{\gamma} (b)^n \big) ~.
\end{equation}
First, we have holomorphic Wilson lines.
These can be understood as coming from standard Wilson lines in 4d CS that are forced to occupy special curves in spacetime.
Second, we have the ad-invariant polynomials of $b$, which can be understood as coming from the higher spin local charges.
Reducing to 1d, only the second set of conserved charges survive.

The Hamiltonian analysis of 4d CS presented in~\cite{Vicedo:2019dej} relates 4d CS to affine Gaudin models.
The Poisson bracket of the Lax matrix is ultralocal for order defects and non-ultralocal for disorder defects.
Similarly, the Hamiltonian formulation of 3d BF~\cite{Vicedo:2022mrm} yields finite Gaudin models with a Poisson bracket satisfying the Lax algebra.
The Hamiltonian analysis of 4d hBF \cite{Winstone:2023fpe} shows that, even in a disorder defect setup, the Poisson bracket of the associated 2d theories is always ultralocal.

The above summary indicates that 2d holomorphic integrable systems sit somewhere between the more typical and well-studied 1d and 2d integrable systems.

\medskip

When constructing 3d and 4d integrable systems from 5d CS and 6d hCS, their integrability will be partially holomorphic and holomorphic respectively.
The underlying complex manifold of 6d hCS is often~\cite{Bittleston:2020hfv,Costello:2021bah,Cole:2023umd,Cole:2024sje} taken to be twistor space $\mathbb{PT} = \mathsf{Tot} \big( \cO(1) \oplus \cO(1) \to \CP^1 \big)$, and the underlying complex manifold of 5d CS is often~\cite{Popov:2005uv,Adamo:2017xaf,Bittleston:2020hfv} taken to be minitwistor space $\mathbb{MT} = \mathsf{Tot} \big( \cO(2) \to \CP^1 \big)$.
This is similar to the underlying complex manifold $\mathsf{Tot} \big( \cO(1) \to \CP^1 \big)$ that we took for 4d hBF in this paper, and leads to the common feature of explicit $z$-dependence in the Lax equation.
Moreover, in all of these theories holomorphic Wilson lines are amongst the natural operators that are expected to play a role in the integrability of the associated field theories, see \cite{Kmec:2025ftx} for example.

While there are similarities between the 2d IFTs we find from 4d hBF and the 3d and 4d IFTs constructed from 5d CS and 6d hCS, there are also differences.
In particular, the minimal setup in 5d CS and 6d hCS gives a one-field model with an invertible kinetic term~\cite{Bittleston:2020hfv}.
By comparison, the minimal non-trivial setup in 4d hBF gives a two-field model with a non-invertible kinetic term.
Therefore, some of the technical problems with quantization of the 2d holomorphic integrable theories constructed in this paper will not apply to their higher-dimensional counterparts.

Despite these differences, there is a further avenue by which our results might be used to learn something about higher-dimensional integrable models.
Higher group Chern-Simons theories, including 5d 2-CS, have recently been proposed as an alternative method for constructing higher-dimensional integrable theories \cite{Schenkel:2024dcd,Chen:2024axr}.
Supposing that there are multiple valid notions of integrability in more than two dimensions, it would be interesting to explore if they are related in a similar way to those that we have explored in this paper.
We could also consider holomorphic versions of these theories, such as 8d holomorphic 2-CS and reductions thereof, constructing a similar network of theories to the one we have investigated.
This may give rise to further alternative ``higher'' notions of integrability in 2d originating from 4d 2-BF theory.
\unskip\footnote{We thank B.\ Vicedo for this suggestion.}

\medskip

Let us now turn to the 2d IFT~\eqref{eq:2dift} that we constructed from 4d hBF.
Solving the equations of motion, we found that the solution depended on a free function of spacetime.
This indicates the presence of a gauge symmetry, also evident in the Hamiltonian analysis, raising the question of how to quantize the classical theory.
There are a number of different approaches.
First, we could attempt a direct analysis, for example, using the BV-BRST formalism to analyse the gauge symmetry in the Lagrangian formalism.
Second, assuming that the chiral symmetries survive quantization, we could take the factorisation~\eqref{eq:factorisation} to hold in the quantum theory and quantize the component fields, similar in spirit to the quantization of the 2d WZW model~\cite{Witten:1983ar,Gawedzki:2001rm}.
Finally, we could treat 4d CS as a regularisation of 4d hBF, first quantizing the 2d IFT~\eqref{eq:4dcssimple} found from 4d CS, and then taking the $k\to 0$ limit.
This is analogous to the expectation that 3d CS at the critical level describes the quantization of 3d BF~\cite{Aganagic:2017tvx,Gaiotto:2021tsq,Zeng:2021zef}.

Even if we managed to quantize these theories, this does not necessarily mean that they are quantum integrable.
To determine this requires an analysis of the quantum symmetries of the theory.
Rather than looking at the quantum symmetries of the 2d IFT, we could instead ask whether or not 4d hBF has a gauge anomaly,\footnote{We thank R.\ Bittleston for discussions on this point.} which would obstruct its quantization.
The gauge anomaly in the higher-dimensional holomorphic-topological theory is interpreted as an obstruction to the quantum integrability of the lower-dimensional IFT~\cite{Costello:2021bah,Bittleston:2022nfr}.
In the case that there is an anomaly, it can sometimes be cancelled by coupling the holomorphic-topological theory to additional matter.
This leads to new degrees of freedom in the lower-dimensional IFT repairing the quantum integrability.

\medskip

Another direction for future research is to modify the defining data of 4d hBF and explore the class of associated 2d IFTs.
In the context of 4d CS, different meromorphic $(1,0)$-forms and different boundary conditions are known to lead to different 2d IFTs, some of which are related to one another by deformations or dualities.
The analogous setups could be studied in the context of 4d hBF, potentially establishing families of 2d IFTs with continuous deformation parameters.

\medskip

Finally, our main theory of interest, 4d hBF, is also of interest in the context of supersymmetric gauge theories.
Holomorphic-topological gauge theories can be constructed as twists of supersymmetric gauge theories where they can be used to study protected sectors of operators in the full supersymmetric theory.
For example, 4d hBF was studied in \cite{Budzik:2023xbr} as the holomorphic twist of $\mathcal{N} = 1$ supersymmetric Yang-Mills where it captures properties of quarter-BPS operators.
It would be interesting to make further connections between these two perspectives on holomorphic-topological gauge theories.

\section*{Acknowledgements}

We would like to thank Nicola Baglioni, Daniele Bielli, Roland Bittleston, Kasia Budzik, Vincent Caudrelier, Kevin Costello, Ryan Cullinan, Sibylle Driezen, Christian Ferko, Sylvain Lacroix, Joaquin Liniado, Adrian Lopez-Raven, Jamie Pearson, Daniel Thompson, Benoit Vicedo and Masahito Yamazaki for interesting discussions.
\\[0.15cm]\noindent
LTC was supported by a UKRI Future Leaders Fellowship (grant number MR/T018909/1) and by an ERC Consolidator/UKRI Frontier grant (TwistorQFT EP/Z000157/1).
BH was supported by a UKRI Future Leaders Fellowship (grant number MR/T018909/1) and by a UKRI Future Leaders Fellowship Renewal (grant number UKRI2067).
We would also like to thank the organisers of the \textit{Workshop on Higher-d Integrability} in Favignana, Italy where part of this work was completed.

\begin{bibtex}[\jobname]

@article{Vicedo:2019dej,
author = "Vicedo, Benoit",
title = "{4D Chern\textendash{}Simons theory and affine Gaudin models}",
eprint = "1908.07511",
archivePrefix = "arXiv",
primaryClass = "hep-th",
doi = "10.1007/s11005-021-01354-9",
journal = "Lett. Math. Phys.",
volume = "111",
number = "1",
pages = "24",
year = "2021"
}

@article{Vicedo:2022mrm,
author = "Vicedo, Benoit and Winstone, Jennifer",
title = "{3-Dimensional mixed BF theory and Hitchin\textquoteright{}s integrable system}",
eprint = "2201.07300",
archivePrefix = "arXiv",
primaryClass = "hep-th",
doi = "10.1007/s11005-022-01567-6",
journal = "Lett. Math. Phys.",
volume = "112",
number = "4",
pages = "79",
year = "2022"
}

@article{Bittleston:2022nfr,
author = "Bittleston, Roland and Skinner, David and Sharma, Atul",
title = "{Quantizing the Non-linear Graviton}",
eprint = "2208.12701",
archivePrefix = "arXiv",
primaryClass = "hep-th",
doi = "10.1007/s00220-023-04828-0",
journal = "Commun. Math. Phys.",
volume = "403",
number = "3",
pages = "1543--1609",
year = "2023"
}

@article{Budzik:2023xbr,
author = "Budzik, Kasia and Gaiotto, Davide and Kulp, Justin and Williams, Brian R. and Wu, Jingxiang and Yu, Matthew",
title = "{Semi-chiral operators in 4d $ \mathcal{N} $ = 1 gauge theories}",
eprint = "2306.01039",
archivePrefix = "arXiv",
primaryClass = "hep-th",
doi = "10.1007/JHEP05(2024)245",
journal = "JHEP",
volume = "05",
pages = "245",
year = "2024"
}

@article{Bittleston:2020hfv,
author = "Bittleston, Roland and Skinner, David",
title = "{Twistors, the ASD Yang-Mills equations and 4d Chern-Simons theory}",
eprint = "2011.04638",
archivePrefix = "arXiv",
primaryClass = "hep-th",
doi = "10.1007/JHEP02(2023)227",
journal = "JHEP",
volume = "02",
pages = "227",
year = "2023"
}

@article{Bittleston:2025gxr,
author = "Bittleston, Roland and Heuveline, Simon and Raghavendran, Surya and Skinner, David",
title = "{Non-Commutative Gauge Theory at the Beach}",
eprint = "2509.20643",
archivePrefix = "arXiv",
primaryClass = "hep-th",
month = "9",
year = "2025"
}

@article{Costello:2013zra,
author = "Costello, Kevin",
title = "{Supersymmetric gauge theory and the Yangian}",
eprint = "1303.2632",
archivePrefix = "arXiv",
primaryClass = "hep-th",
month = "3",
year = "2013"
}

@article{Costello:2013sla,
author = "Costello, Kevin",
editor = "Donagi, Ron and Douglas, Michael R. and Kamenova, Ljudmila and Rocek, Martin",
title = "{Integrable lattice models from four-dimensional field theories}",
eprint = "1308.0370",
archivePrefix = "arXiv",
primaryClass = "hep-th",
doi = "10.1090/pspum/088/01483",
journal = "Proc. Symp. Pure Math.",
volume = "88",
pages = "3--24",
year = "2014"
}

@article{Costello:2017dso,
author = "Costello, Kevin and Witten, Edward and Yamazaki, Masahito",
title = "{Gauge Theory and Integrability, I}",
eprint = "1709.09993",
archivePrefix = "arXiv",
primaryClass = "hep-th",
reportNumber = "IPMU17-0136",
doi = "10.4310/ICCM.2018.v6.n1.a6",
journal = "ICCM Not.",
volume = "06",
number = "1",
pages = "46--119",
year = "2018"
}

@article{Costello:2018gyb,
author = "Costello, Kevin and Witten, Edward and Yamazaki, Masahito",
title = "{Gauge Theory and Integrability, II}",
eprint = "1802.01579",
archivePrefix = "arXiv",
primaryClass = "hep-th",
reportNumber = "IPMU18-0025",
doi = "10.4310/ICCM.2018.v6.n1.a7",
journal = "ICCM Not.",
volume = "06",
number = "1",
pages = "120--146",
year = "2018"
}

@article{Costello:2019tri,
author = "Costello, Kevin and Yamazaki, Masahito",
title = "{Gauge Theory And Integrability, III}",
eprint = "1908.02289",
archivePrefix = "arXiv",
primaryClass = "hep-th",
reportNumber = "IPMU19-0110",
month = "8",
year = "2019"
}

@article{Cole:2024sje,
author = "Cole, Lewis T. and Cullinan, Ryan A. and Hoare, Ben and Liniado, Joaquin and Thompson, Daniel C.",
title = "{Gauging the diamond: integrable coset models from twistor space}",
eprint = "2407.09479",
archivePrefix = "arXiv",
primaryClass = "hep-th",
doi = "10.1007/JHEP12(2024)202",
journal = "JHEP",
volume = "12",
pages = "202",
year = "2024"
}

@article{Cole:2023umd,
author = "Cole, Lewis T. and Cullinan, Ryan A. and Hoare, Ben and Liniado, Joaquin and Thompson, Daniel C.",
title = "{Integrable deformations from twistor space}",
eprint = "2311.17551",
archivePrefix = "arXiv",
primaryClass = "hep-th",
doi = "10.21468/SciPostPhys.17.1.008",
journal = "SciPost Phys.",
volume = "17",
number = "1",
pages = "008",
year = "2024"
}

@article{Caudrelier:2025xtx,
author = "Caudrelier, Vincent and Harland, Derek and Singh, Anup Anand and Vicedo, Benoit",
title = "{The 3d mixed BF Lagrangian 1-form: a variational formulation of Hitchin's integrable system}",
eprint = "2509.05127",
archivePrefix = "arXiv",
primaryClass = "math-ph",
month = "9",
year = "2025"
}

@phdthesis{Winstone:2023fpe,
author = "Winstone, Jennifer",
title = "{Aspects of Finite Gaudin Models: Separation of Variables and Description from 3dBF Theory}",
school = "Dept. Math., York U., England",
year = "2023"
}

@article{Lacroix:2021iit,
author = "Lacroix, Sylvain",
title = "{Four-dimensional Chern{\textendash}Simons theory and integrable field theories}",
eprint = "2109.14278",
archivePrefix = "arXiv",
primaryClass = "hep-th",
doi = "10.1088/1751-8121/ac48ed",
journal = "J. Phys. A",
volume = "55",
number = "8",
pages = "083001",
year = "2022"
}

@article{Costello:2021bah,
author = "Costello, Kevin J.",
title = "{Quantizing local holomorphic field theories on twistor space}",
eprint = "2111.08879",
archivePrefix = "arXiv",
primaryClass = "hep-th",
month = "11",
year = "2021"
}

@talk{Costello:talk,
author = {Costello, Kevin},
title = {Topological strings, twistors and Skyrmions},
publisher = {The Western Hemisphere Colloquium on Geometry and Physics},
URL = {https://www.youtube.com/watch?v=ZlDNpPHvA8A},
year = {2020},
note = ""
}

@article{Benini:2020skc,
author = "Benini, Marco and Schenkel, Alexander and Vicedo, Benoit",
title = "{Homotopical Analysis of 4d Chern-Simons Theory and Integrable Field Theories}",
eprint = "2008.01829",
archivePrefix = "arXiv",
primaryClass = "hep-th",
doi = "10.1007/s00220-021-04304-7",
journal = "Commun. Math. Phys.",
volume = "389",
number = "3",
pages = "1417--1443",
year = "2022"
}

@article{Delduc:2019whp,
author = "Delduc, Francois and Lacroix, Sylvain and Magro, Marc and Vicedo, Benoit",
title = "{A unifying 2D action for integrable $\sigma $-models from 4D Chern{\textendash}Simons theory}",
eprint = "1909.13824",
archivePrefix = "arXiv",
primaryClass = "hep-th",
doi = "10.1007/s11005-020-01268-y",
journal = "Lett. Math. Phys.",
volume = "110",
number = "7",
pages = "1645--1687",
year = "2020"
}

@article{Lacroix:2017isl,
author = "Lacroix, Sylvain and Magro, Marc and Vicedo, Benoit",
title = "{Local charges in involution and hierarchies in integrable sigma-models}",
eprint = "1703.01951",
archivePrefix = "arXiv",
primaryClass = "hep-th",
doi = "10.1007/JHEP09(2017)117",
journal = "JHEP",
volume = "09",
pages = "117",
year = "2017"
}

@article{Maillet:1985fn,
author = "Maillet, Jean Michel",
title = "{Kac-moody Algebra and Extended {Yang-Baxter} Relations in the O($N$) Nonlinear $\sigma$ Model}",
reportNumber = "PAR LPTHE 85-22",
doi = "10.1016/0370-2693(85)91075-5",
journal = "Phys. Lett. B",
volume = "162",
pages = "137--142",
year = "1985"
}

@article{Maillet:1985ek,
author = "Maillet, Jean Michel",
title = "{New Integrable Canonical Structures in Two-dimensional Models}",
reportNumber = "PAR/LPTHE-85-32",
doi = "10.1016/0550-3213(86)90365-2",
journal = "Nucl. Phys. B",
volume = "269",
pages = "54--76",
year = "1986"
}

@article{Ashwinkumar:2023zbu,
author = "Ashwinkumar, Meer and Sakamoto, Jun-ichi and Yamazaki, Masahito",
title = "{Dualities and Discretizations of Integrable Quantum Field Theories from 4d Chern-Simons Theory}",
eprint = "2309.14412",
archivePrefix = "arXiv",
primaryClass = "hep-th",
doi = "10.4310/ATMP.251118001759",
journal = "Adv. Theor. Math. Phys.",
volume = "29",
pages = "1509--1694",
year = "2025"
}

@article{Adamo:2017xaf,
author = "Adamo, Tim and Skinner, David and Williams, Jack",
title = "{Minitwistors and 3d Yang-Mills-Higgs theory}",
eprint = "1712.09604",
archivePrefix = "arXiv",
primaryClass = "hep-th",
reportNumber = "IMPERIAL-TP-TA-2017-04",
doi = "10.1063/1.5030417",
journal = "J. Math. Phys.",
volume = "59",
number = "12",
pages = "122301",
year = "2018"
}

@article{Popov:2005uv,
author = "Popov, Alexander D. and Saemann, Christian and Wolf, Martin",
title = "{The Topological B-model on a mini-supertwistor space and supersymmetric Bogomolny monopole equations}",
eprint = "hep-th/0505161",
archivePrefix = "arXiv",
reportNumber = "ITP-UH-07-05",
doi = "10.1088/1126-6708/2005/10/058",
journal = "JHEP",
volume = "10",
pages = "058",
year = "2005"
}

@article{Lax:1968gpe,
author = "Lax, Peter D.",
title = "{Integrals of nonlinear equations of evolution and solitary waves}",
doi = "10.1002/cpa.3160210503",
journal = "Commun. Pure Appl. Math.",
volume = "21",
number = "5",
pages = "467--490",
year = "1968"
}

@book{FaddeevTakhtajan,
author = "L. Faddeev and L. L.A. Takhtajan",
title = "Hamiltonian methods in the theory of solitons",
publisher = "Springer",
year = "1987",
}

@book{BabelonBernardTalon,
author = "O. Babelon, D. Bernard and M. Talon",
title = "Introduction to Classical Integrable Models",
publisher = "Cambridge University Press",
year = "2003"
}

@article{Leznov:1986mx,
author = "Leznov, A. N. and Mukhtarov, M. A.",
title = "{Deformation of Algebras and Solution of Selfduality Equation}",
doi = "10.1063/1.527748",
journal = "J. Math. Phys.",
volume = "28",
pages = "2574--2578",
year = "1987"
}

@article{Parkes:1992rz,
author = "Parkes, Andrew",
title = "{A Cubic action for selfdual Yang-Mills}",
eprint = "hep-th/9203074",
archivePrefix = "arXiv",
reportNumber = "ETH-TH-92-14",
doi = "10.1016/0370-2693(92)91773-3",
journal = "Phys. Lett. B",
volume = "286",
pages = "265--270",
year = "1992"
}

@article{Losev:1995cr,
author = "Losev, Andrei and Moore, Gregory W. and Nekrasov, Nikita and Shatashvili, Samson",
editor = "Bars, I. and Bouwknegt, P. and Minahan, J. and Nemeschansky, D. and Pilch, K. and Saleur, H. and Warner, N. P.",
title = "{Four-Dimensional Avatars of Two-Dimensional RCFT}",
eprint = "hep-th/9509151",
archivePrefix = "arXiv",
reportNumber = "PUPT-1564, ITEP-5-95, YCTP-P-15-95",
doi = "10.1016/0920-5632(96)00015-1",
journal = "Nucl. Phys. B Proc. Suppl.",
volume = "46",
pages = "130--145",
year = "1996"
}

@phdthesis{NekrasovThesis,
author = {Nekrasov, N.},
year = {1996},
month = {06},
pages = {},
title = {Four dimensional holomorphic theories},
doi = {10.13140/RG.2.1.4116.2481}
}

@article{Sklyanin:1982tf,
author = "Sklyanin, E. K.",
title = "{Some algebraic structures connected with the Yang-Baxter equation}",
doi = "10.1007/BF01077848",
journal = "Funct. Anal. Appl.",
volume = "16",
pages = "263--270",
year = "1982"
}

@article{Ward:1985gz,
author = "Ward, R. S.",
title = "{Integrable and solvable systems, and relations among them}",
doi = "10.1098/rsta.1985.0051",
journal = "Phil. Trans. Roy. Soc. Lond. A",
volume = "315",
pages = "451--457",
year = "1985"
}

@book{mason1996integrability,
title={Integrability, self-duality, and twistor theory},
author={Mason, Lionel J and Woodhouse, Nicholas Michael John},
number={15},
year={1996},
publisher={Oxford University Press}
}

@article{Schenkel:2024dcd,
author = "Schenkel, Alexander and Vicedo, Beno{\textasciicircum}it",
title = "{5d 2-Chern-Simons Theory and 3d Integrable Field Theories}",
eprint = "2405.08083",
archivePrefix = "arXiv",
primaryClass = "hep-th",
doi = "10.1007/s00220-024-05170-9",
journal = "Commun. Math. Phys.",
volume = "405",
number = "12",
pages = "293",
year = "2024"
}

@article{Chen:2024axr,
author = "Chen, Hank and Liniado, Joaquin",
title = "{Higher gauge theory and integrability}",
eprint = "2405.18625",
archivePrefix = "arXiv",
primaryClass = "hep-th",
doi = "10.1103/PhysRevD.110.086017",
journal = "Phys. Rev. D",
volume = "110",
number = "8",
pages = "086017",
year = "2024"
}

@article{Aganagic:2017tvx,
author = "Aganagic, Mina and Costello, Kevin and McNamara, Jacob and Vafa, Cumrun",
title = "{Topological Chern-Simons/Matter Theories}",
eprint = "1706.09977",
archivePrefix = "arXiv",
primaryClass = "hep-th",
month = "6",
year = "2017"
}

@article{Gaiotto:2021tsq,
author = "Gaiotto, Davide and Witten, Edward",
title = "{Gauge Theory and the Analytic Form of the Geometric Langlands Program}",
eprint = "2107.01732",
archivePrefix = "arXiv",
primaryClass = "hep-th",
doi = "10.1007/s00023-022-01225-6",
journal = "Annales Henri Poincare",
volume = "25",
number = "1",
pages = "557--671",
year = "2024"
}

@article{Zeng:2021zef,
author = "Zeng, Keyou",
title = "{Monopole operators and bulk-boundary relation in holomorphic topological theories}",
eprint = "2111.00955",
archivePrefix = "arXiv",
primaryClass = "hep-th",
doi = "10.21468/SciPostPhys.14.6.153",
journal = "SciPost Phys.",
volume = "14",
number = "6",
pages = "153",
year = "2023"
}

@article{Witten:1983ar,
author = "Witten, Edward",
editor = "Stone, M.",
title = "{Nonabelian Bosonization in Two-Dimensions}",
reportNumber = "PRINT-83-0934 (PRINCETON)",
doi = "10.1007/BF01215276",
journal = "Commun. Math. Phys.",
volume = "92",
pages = "455--472",
year = "1984"
}

@article{Gawedzki:2001rm,
author = "Gawedzki, Krzysztof and Todorov, Ivan and Tran-Ngoc-Bich, Pascal",
title = "{Canonical quantization of the boundary Wess-Zumino-Witten model}",
eprint = "hep-th/0101170",
archivePrefix = "arXiv",
doi = "10.1007/s00220-004-1107-6",
journal = "Commun. Math. Phys.",
volume = "248",
pages = "217--254",
year = "2004"
}

@article{Kmec:2025ftx,
author = "Kmec, Adam and Mason, Lionel and Ruzziconi, Romain and Sharma, Atul",
title = "{S-algebra in gauge theory: twistor, spacetime and holographic perspectives}",
eprint = "2506.01888",
archivePrefix = "arXiv",
primaryClass = "hep-th",
doi = "10.1088/1361-6382/ae0673",
journal = "Class. Quant. Grav.",
volume = "42",
number = "19",
pages = "195008",
year = "2025"
}

@article{Yang:1977zf,
author = "Yang, Chen Ning",
title = "{Condition of Selfduality for SU(2) Gauge Fields on Euclidean Four-Dimensional Space}",
reportNumber = "ITP-SB-77-33",
doi = "10.1103/PhysRevLett.38.1377",
journal = "Phys. Rev. Lett.",
volume = "38",
pages = "1377",
year = "1977"
}

@article{Pohlmeyer:1979ya,
author = "Pohlmeyer, K.",
title = "{On the Lagrangian Theory of Anti(self)dual Fields in Four-dimensional Euclidean Space}",
reportNumber = "Print-79-0680 (FREIBURG)",
doi = "10.1007/BF01200109",
journal = "Commun. Math. Phys.",
volume = "72",
pages = "37",
year = "1980"
}

@article{donaldson1985anti,
title = {Anti self-dual Yang-Mills connections over complex algebraic surfaces and stable vector bundles},
author = {Donaldson, Simon K},
doi = "10.1112/plms/s3-50.1.1",
journal = {Proceedings of the London Mathematical Society},
volume = {3},
number = {1},
pages = {1--26},
year = {1985},
publisher = {Wiley Online Library}
}

@article{Newman:1978ze,
author = "Newman, E. T.",
title = "{Source-Free Yang-Mills Theories}",
doi = "10.1103/PhysRevD.18.2901",
journal = "Phys. Rev. D",
volume = "18",
pages = "2901--2908",
year = "1978"
}

@article{Leznov:1986up,
author = "Leznov, A. N.",
title = "{On Equivalence of Four-dimensional Selfduality Equations to Continual Analog of the Main Chiral Field Problem. (In Russian)}",
reportNumber = "IFVE-86-188",
doi = "10.1007/BF01017594",
journal = "Teor. Mat. Fiz.",
volume = "73",
pages = "302--307",
year = "1987"
}

@article{Witten:2003nn,
author = "Witten, Edward",
title = "{Perturbative gauge theory as a string theory in twistor space}",
eprint = "hep-th/0312171",
archivePrefix = "arXiv",
doi = "10.1007/s00220-004-1187-3",
journal = "Commun. Math. Phys.",
volume = "252",
pages = "189--258",
year = "2004"
}

@article{Mason:2005zm,
author = "Mason, L. J.",
title = "{Twistor actions for non-self-dual fields: A Derivation of twistor-string theory}",
eprint = "hep-th/0507269",
archivePrefix = "arXiv",
doi = "10.1088/1126-6708/2005/10/009",
journal = "JHEP",
volume = "10",
pages = "009",
year = "2005"
}

@article{Vicedo:2017cge,
author = "Vicedo, Benoit",
title = "{On integrable field theories as dihedral affine Gaudin models}",
eprint = "1701.04856",
archivePrefix = "arXiv",
primaryClass = "hep-th",
doi = "10.1093/imrn/rny128",
journal = "Int. Math. Res. Not.",
volume = "2020",
number = "15",
pages = "4513--4601",
year = "2020"
}

@article{Delduc:2012qb,
author = "Delduc, F. and Magro, M. and Vicedo, B.",
title = "{Alleviating the non-ultralocality of coset sigma models through a generalized Faddeev-Reshetikhin procedure}",
eprint = "1204.0766",
archivePrefix = "arXiv",
primaryClass = "hep-th",
doi = "10.1007/JHEP08(2012)019",
journal = "JHEP",
volume = "08",
pages = "019",
year = "2012"
}

@article{Delduc:2023exz,
author = "Delduc, Francois and Hoare, Ben and Magro, Marc",
title = "{Towards a quadratic Poisson algebra for the subtracted classical monodromy of symmetric space sine-Gordon theories}",
eprint = "2309.15722",
archivePrefix = "arXiv",
primaryClass = "hep-th",
doi = "10.1088/1751-8121/ad1d91",
journal = "J. Phys. A",
volume = "57",
number = "6",
pages = "065401",
year = "2024"
}

@article{Takhtajan:1979iv,
author = "Takhtajan, L. A. and Faddeev, L. D.",
title = "{The Quantum method of the inverse problem and the Heisenberg XYZ model}",
journal = "Russ. Math. Surveys",
volume = "34",
number = "5",
pages = "11--68",
year = "1979"
}

@article{Kulish:1979if,
author = "Kulish, P. P. and Sklyanin, E. K.",
title = "{Quantum inverse scattering method and the Heisenberg ferromagnet}",
doi = "10.1016/0375-9601(79)90365-7",
journal = "Phys. Lett. A",
volume = "70",
pages = "461--463",
year = "1979"
}

@article{Sklyanin:1979pfu,
author = "Sklyanin, E. K. and Takhtadzhyan, L. A. and Faddeev, L. D.",
title = "{Quantum inverse problem method. I}",
reportNumber = "LOMI-P-1-79",
doi = "10.1007/BF01018718",
journal = "Theor. Math. Phys.",
volume = "40",
number = "2",
pages = "688--706",
year = "1979"
}

@article{Bykov:2021dbk,
author = "Bykov, Dmitri",
title = "{Sigma models as Gross{\textendash}Neveu models}",
eprint = "2106.15598",
archivePrefix = "arXiv",
primaryClass = "hep-th",
doi = "10.1134/S0040577921080018",
journal = "Teor. Mat. Fiz.",
volume = "208",
number = "2",
pages = "165--179",
year = "2021"
}

@article{Delduc:2019lpe,
author = "Delduc, Francois and Kameyama, Takashi and Lacroix, Sylvain and Magro, Marc and Vicedo, Benoit",
title = "{Ultralocal Lax connection for para-complex $\mathbb{Z}_T$-cosets}",
eprint = "1909.00742",
archivePrefix = "arXiv",
primaryClass = "hep-th",
doi = "10.1016/j.nuclphysb.2019.114821",
journal = "Nucl. Phys. B",
volume = "949",
pages = "114821",
year = "2019"
}

@article{Semenov-Tian-Shansky:1995zdv,
author = "Semenov-Tian-Shansky, Michael and Sevostyanov, Alexei",
title = "{Classical and quantum nonultralocal systems on the lattice}",
eprint = "hep-th/9509029",
archivePrefix = "arXiv",
month = "9",
year = "1995"
}

@article{Brodbeck:1999ib,
author = "Brodbeck, Othmar and Zagermann, Marco",
title = "{Dimensionally reduced gravity, Hermitian symmetric spaces and the Ashtekar variables}",
eprint = "gr-qc/9911118",
archivePrefix = "arXiv",
reportNumber = "PSU-TH-223",
doi = "10.1088/0264-9381/17/14/310",
journal = "Class. Quant. Grav.",
volume = "17",
pages = "2749--2764",
year = "2000"
}

@article{Faddeev:1985qu,
author = "Faddeev, L. D. and Reshetikhin, N. Yu.",
title = "{Integrability of the Principal Chiral Field Model in (1+1)-dimension}",
reportNumber = "LOMI-E-2-85",
doi = "10.1016/0003-4916(86)90201-0",
journal = "Annals Phys.",
volume = "167",
pages = "227",
year = "1986"
}

@article{Yamazaki:2019prm,
author = "Yamazaki, Masahito",
title = "{New T-duality for Chern-Simons Theory}",
eprint = "1904.04976",
archivePrefix = "arXiv",
primaryClass = "hep-th",
reportNumber = "IPMU19-0049",
doi = "10.1007/JHEP12(2019)090",
journal = "JHEP",
volume = "12",
pages = "090",
year = "2019"
}

@article{Ferko:2024ali,
author = "Ferko, Christian and Smith, Liam",
title = "{Infinite Family of Integrable Sigma Models Using Auxiliary Fields}",
eprint = "2405.05899",
archivePrefix = "arXiv",
primaryClass = "hep-th",
doi = "10.1103/PhysRevLett.133.131602",
journal = "Phys. Rev. Lett.",
volume = "133",
number = "13",
pages = "131602",
year = "2024"
}

@article{Bielli:2024ach,
author = "Bielli, Daniele and Ferko, Christian and Smith, Liam and Tartaglino-Mazzucchelli, Gabriele",
title = "{Integrable higher-spin deformations of sigma models from auxiliary fields}",
eprint = "2407.16338",
archivePrefix = "arXiv",
primaryClass = "hep-th",
doi = "10.1103/PhysRevD.111.066010",
journal = "Phys. Rev. D",
volume = "111",
number = "6",
pages = "066010",
year = "2025"
}

@article{Sakamoto:2025hwi,
author = "Sakamoto, Jun-ichi and Tateo, Roberto and Yamazaki, Masahito",
title = "{$T\bar{T}$ and root-$T\bar{T}$ deformations in four-dimensional Chern-Simons theory}",
eprint = "2509.12303",
archivePrefix = "arXiv",
primaryClass = "hep-th",
month = "9",
year = "2025"
}

@article{Fukushima:2024nxm,
author = "Fukushima, Osamu and Yoshida, Kentaroh",
title = "{4D Chern-Simons theory with auxiliary fields}",
eprint = "2407.02204",
archivePrefix = "arXiv",
primaryClass = "hep-th",
reportNumber = "RIKEN-iTHEMS-Report-24, STUPP-24-270",
doi = "10.1007/JHEP09(2025)001",
journal = "JHEP",
volume = "09",
pages = "001",
year = "2025"
}

@article{Babaei-Aghbolagh:2025uoz,
author = "Babaei-Aghbolagh, H. and Chen, Bin and He, Song",
title = "{Root-$T\bar{T}$ flows unify 4D duality-invariant electrodynamics and 2D integrable sigma models}",
eprint = "2507.22808",
archivePrefix = "arXiv",
primaryClass = "hep-th",
doi = "10.1103/1r4p-3r5q",
journal = "Phys. Rev. D",
volume = "112",
number = "10",
pages = "L101702",
year = "2025"
}

@article{Babaei-Aghbolagh:2025hlm,
author = "Babaei-Aghbolagh, H. and Chen, Bin and He, Song",
title = "{Integrable Sigma Models and Universal Root $T\bar{T}$ Deformation via Courant-Hilbert Approach}",
eprint = "2509.17075",
archivePrefix = "arXiv",
primaryClass = "hep-th",
month = "9",
year = "2025"
}

@article{Baglioni:2025tsc,
author = "Baglioni, Nicola and Bielli, Daniele and Galli, Michele and Tartaglino-Mazzucchelli, Gabriele",
title = "{Relating auxiliary field formulations of $4d$ duality-invariant and $2d$ integrable field theories}",
eprint = "2512.21982",
archivePrefix = "arXiv",
primaryClass = "hep-th",
month = "12",
year = "2025"
}

@article{Fukushima:2025tlj,
author = "Fukushima, Osamu and Matsumoto, Takaki and Yoshida, Kentaroh",
title = "{The Courant-Hilbert construction in 4D Chern-Simons theory}",
eprint = "2509.22080",
archivePrefix = "arXiv",
primaryClass = "hep-th",
reportNumber = "RIKEN-iTHEMS-Report-25, STUPP-25-288",
month = "9",
year = "2025"
}

@article{Moosavian:2025nyv,
author = "Moosavian, Seyed Faroogh and Yamazaki, Masahito and Zhou, Yehao",
title = "{Hyperbolic Monopoles, (Semi-)Holomorphic Chern-Simons Theories, and Generalized Chiral Potts Models}",
eprint = "2502.17545",
archivePrefix = "arXiv",
primaryClass = "hep-th",
month = "2",
year = "2025"
}

@article{Popov:1999cq,
author = "Popov, Alexander D.",
title = "{Holomorphic analogs of topological gauge theories}",
eprint = "hep-th/9909135",
archivePrefix = "arXiv",
doi = "10.1016/S0370-2693(99)01464-1",
journal = "Phys. Lett. B",
volume = "473",
pages = "65--72",
year = "2000"
}

@article{Costello:2016nkh,
author = "Costello, Kevin",
title = "{M-theory in the Omega-background and 5-dimensional non-commutative gauge theory}",
eprint = "1610.04144",
archivePrefix = "arXiv",
primaryClass = "hep-th",
month = "10",
year = "2016"
}

@article{Caudrelier:2020xtn,
    author = "Caudrelier, Vincent and Stoppato, Matteo and Vicedo, Benoit",
    title = "{On the Zakharov{\textendash}Mikhailov action: $4\hbox {d}$ Chern{\textendash}Simons origin and covariant Poisson algebra of the Lax connection}",
    eprint = "2012.04431",
    archivePrefix = "arXiv",
    primaryClass = "hep-th",
    doi = "10.1007/s11005-021-01425-x",
    journal = "Lett. Math. Phys.",
    volume = "111",
    number = "3",
    pages = "82",
    year = "2021"
}

@article{Fukushima:2020tqv,
    author = "Fukushima, Osamu and Sakamoto, Jun-Ichi and Yoshida, Kentaroh",
    title = "{Faddeev-Reshetikhin model from a 4D Chern-Simons theory}",
    eprint = "2012.07370",
    archivePrefix = "arXiv",
    primaryClass = "hep-th",
    reportNumber = "KUNS-2847",
    doi = "10.1007/JHEP02(2021)115",
    journal = "JHEP",
    volume = "02",
    pages = "115",
    year = "2021"
}

\end{bibtex}

\bibliographystyle{nb}
\bibliography{\jobname}

\end{document}